\documentstyle[12pt]{article}

\input epsf

\begin{document}
\begin{titlepage}
\today          \hfill
\begin{center}
\hfill    OITS-682 \\

\vskip .05in

{\large Can multi-TeV (top and other) squarks be natural in
gauge mediation?}
\footnote{This work is supported by DOE Grant DE-FG03-96ER40969.}
\vskip .15in
Kaustubh Agashe \footnote{email: agashe@oregon.uoregon.edu}
\vskip .1in
{\em
Institute of Theoretical Science \\
5203 University
of Oregon \\
Eugene OR 97403-5203}
\end{center}

\vskip .05in

\begin{abstract}
We investigate whether multi-TeV ($1-3$ TeV) squarks 
can be natural in models of
gauge mediated SUSY breaking. The idea is that for some boundary condition
of the scalar (Higgs and stop) masses, the Higgs (mass)$^2$, evaluated at the
renormalization scale $\sim O(100)$ GeV, is not very sensitive 
to (boundary values of) the scalar masses
(this has been called ``focussing'' in recent literature). 
Then, the stop masses can
be multi-TeV without leading to
fine-tuning in electroweak symmetry breaking.
{\em Minimal} 
gauge mediation does {\em not} lead to this focussing (for all
values of $\tan \beta$ and the messenger scale):
the (boundary value of) the Higgs mass is too small compared to the
stop masses. 
Also, in minimal gauge mediation, the gaugino masses
are of the same order as the scalar masses so that multi-TeV scalars
implies multi-TeV gauginos (especially gluino) leading to fine-tuning.
We discuss ideas to {\em increase} 
the Higgs mass relative to the stop masses 
(so that focussing can be achieved) and
also to {\em suppress} gaugino 
masses relative to scalar masses (or to modify the
gaugino mass relations)
in {\em non-minimal}
models of
gauge mediation -- then multi-TeV (top and other)
squarks can be natural.
Specific models of
gauge mediation
which incorporate
these ideas and thus have squarks (and in some cases,
the gluino) heavier than a TeV 
without
resulting in fine-tuning are also studied and their
collider signals are
contrasted with those of other models which have multi-TeV squarks.
\end{abstract}

\end{titlepage}

\newpage
\renewcommand{\thepage}{\arabic{page}}
\setcounter{page}{1}

\section{Introduction}
In the supersymmetric extension of the SM, the 
quantum corrections to the Higgs (mass)$^2$ cancel between fermions and bosons
in the loops. Thus, SUSY stabilizes the hierarchy between the weak scale and
some high energy scale such as the GUT or the Planck scale.
SUSY can be broken (softly) so that the superpartners of the SM particles 
are heavier than the SM particles (as required by phenomenology), but
the quadratically divergent corrections to the Higgs (mass)$^2$ 
still cancel.
However, with soft SUSY breaking, there are
the logarithmically divergent
corrections to the Higgs (mass)$^2$, proportional to
the soft SUSY breaking masses. Due to the large top 
quark Yukawa coupling, these
radiative corrections
result in 
a negative Higgs (mass)$^2$
(for the Higgs doublet coupled to the top quark), which, in turn,
leads to electroweak symmetry
breaking (EWSB)
without the need (unlike the SM)
to put in 
a bare negative (mass)$^2$ term. 
``by hand''. 
The 
expression for $m_Z$ (at tree-level) is:
\begin{eqnarray}
\frac{1}{2} m_Z^2 & = & - \mu^2 + \frac{m^2_{H_d} - m_{H_u}^2 \tan
^2 \beta}{ \tan ^2 \beta - 1} \nonumber \\
 & \approx & - \mu^2 - m_{H_u}^2 \; (\hbox{for large} \; \tan \beta).
\label{mztree}
\end{eqnarray}
Here, $\mu$ is a supersymmetric mass term for the Higgs,
$\tan \beta$ is the ratio of vev's of the two Higgs doublets
and $m_{H_u}^2$ and
$m_{H_d}^2$ are the soft SUSY breaking masses for the Higgs 
(coupling to the up-type and down-type quarks, respectively)
evaluated at the
weak scale. 

For naturalness of electroweak symmetry
breaking
{\it i.e.,} for the $Z$ mass not to be fine-tuned \footnote{
In general, the $Z$ mass is fine-tuned if it is
{\em too} sensitive to the fundamental parameters of the theory.}
it is necessary that all the terms on the right-side of Eq. (\ref{mztree})
are of order $m_Z^2$, {\it i.e.,} the
Higgs (mass)$^2$ at the weak scale
and $\mu^2$ are  
$O(100 \; \hbox{GeV})^2$.
Typically, this implies that top squarks should be lighter than $\sim 1$ TeV:
heavier top squarks will result 
in Higgs (mass)$^2 \sim  - \hbox{TeV} ^2$ 
(due to renormalization group (RG)
scaling to the weak scale and large top quark Yukawa coupling). 
This will 
necessitate
a large cancellation with $\mu^2$ to obtain the correct value of
$m_Z$ ($\sim 100$ GeV) (see Eq. (\ref{mztree})), 
{\it i.e.},
$m_Z$ will be very sensitive to $\mu$. 
Also, in general,
if the top squark (stop) mass is heavier than $\sim 1$ TeV, then $m_Z$,
in addition to being sensitive to
$\mu$, will
be very sensitive to the {\em stop mass} since it is the stop mass
which results in Higgs (mass)$^2 \sim  - \hbox{TeV} ^2$ at the weak scale.

Recently, Feng, Matchev, Moroi showed that multi-TeV scalars
(in particular top squarks) can be natural
in supergravity mediated models with a {\em specific} boundary condition 
(which includes universal scalar masses)
for
the Higgs and stop masses (at $M_{GUT}$ or $M_{Planck}$)
and for $\tan \beta 
\stackrel{>}{\sim}5$ (with the measured value of 
the top quark mass) \cite{feng}. 
This happens because, for this 
boundary condition, the Higgs (mass)$^2$ evaluated at the renormalization
scale $\sim O(100)$ GeV, is {\em not} 
very sensitive to the boundary value of the
scalar masses (these authors
call this ``focussing''). 
Then, 
multi-TeV
top squarks will {\em not} result in $m ^2 _{H_u} (\hbox{at} \;
\sim 100 \; \hbox{GeV} ) \sim
\;  - \hbox{TeV}^2$
and thus the stop masses can be multi-TeV {\em without} leading to 
fine-tuning in electroweak symmetry breaking. We 
briefly review this idea in section
\ref{focussing}.

In this paper, we study if a similar focussing of (the weak scale
value of) the Higgs (mass)$^2$ can occur in models where SUSY breaking
is mediated to the MSSM at lower energy scales, {\it i.e.}, the
``messenger'' scale is $M_{mess} \ll M_{Planck}$ or $M_{GUT}$.
The only known mediation of SUSY breaking with $M_{mess} \ll M_{Planck}$ 
or $M_{GUT}$ is by (SM or other)
gauge interactions. Gauge mediation (GM)
of SUSY breaking (GMSB) does not
have the supersymmetric
flavor problem (since scalars with the same gauge
quantum numbers are degenerate)
and also has a predictive spectrum,
unlike (generic)
supergravity mediation of SUSY breaking.
So, we analyse whether
in GMSB, the boundary
values of the stop and Higgs masses can be such that the Higgs
(mass)$^2$ at the weak scale is insensitive to the stop masses.
Then, multi-TeV stops (and other squarks)
can be natural in GMSB \footnote{If scalar and gaugino masses are 
{\em sub}-TeV, then (for
moderate and large $\tan \beta$)
electric dipole moments of electron and neutron constrain
the (relative) phases in gaugino mass and $\mu$, $B \mu$ terms to be
less than $O( 10^{-2})$ resulting in fine-tuning. 
With {\em multi}-TeV squarks, $O(0.1)$ or larger
phases might be allowed, thus (partly) alleviating the 
problem of fine-tuning of these phases. This could be a motivation
for multi-TeV scalars in GM: the scalar degeneracy already solves the
supersymmetric flavor problem (even with {\em sub}-TeV scalars).
}.

In minimal GMSB,
a single field which breaks SUSY couples in the superpotential
to ``messengers'' which are vector-like fields in complete
multiplets of $SU(5)$. In this minimal model of
GM (discussed in section \ref{minimal}), the boundary value 
of the Higgs mass is too small compared to the stop mass (since scalar
masses are proportional to (gauge couplings)$^2$, {\it i.e.,}
$\alpha _A$'s) so that focussing does
not occur, {\it i.e.}, the boundary condition necessary
for
the Higgs (mass)$^2$ at the weak scale to be insensitive to the stop
mass requires {\em larger} Higgs mass compared to the stop mass.
Also, in minimal GM, the gaugino masses are comparable
to the scalar masses. So, if the scalar masses are multi-TeV, so will
be the gaugino (especially gluino) masses which will, in turn, result in
large $| m _{H_u} ^2 |$ (through
RG scaling) at the weak scale and hence fine-tuning of the $Z$ mass
\footnote{These facts are well known from the studies of minimal
GM in the literature.}. 

In section \ref{nonminimal}, 
we study how in (non-minimal) models of GMSB the boundary value of the
Higgs mass can be {\em larger} so that the boundary condition for the scalar
masses which results in focussing can be achieved. We also discuss how to
reduce the gaugino masses relative to the scalar masses (or how
to modify the relation between the various gaugino masses) so that,
even if the scalars are multi-TeV, the gaugino masses do not result
in fine-tuning. With focussing
and suppression of gaugino masses (or a modified relation between
the gaugino masses),
multi-TeV ($1-3$ TeV) squarks can be natural in GMSB.

Some specific non-minimal models of GMSB are analysed in
section \ref{specific} which have these
features so that squarks (and in some models the gluino)
can be heavier than a TeV 
without resulting
in fine-tuning in EWSB. We also discuss
the collider signals 
for these models and contrast
these to the signals in minimal
supergravity models with multi-TeV squarks
(and with no fine-tuning due to focussing) \cite{feng}
and also to the signals in the {\em minimal} model of GM with multi-TeV 
squarks
(which {\em is} fine-tuned).

\section{Focussing}
\label{focussing}
We begin with a brief discussion of ``focussing'', {\it i.e.,} the boundary 
condition for the scalar masses such that the Higgs (mass)$^2$ at the weak
scale is insensitive to the scalar masses. The one-loop 
renormalization group equations (RGE's) for the 
(up-type)
Higgs and stop masses, neglecting all Yukawa couplings except the
top quark Yukawa coupling
($\lambda _t$) are \footnote{In section \ref{qlmodel}, we comment on
the contribution of multi-TeV scalar masses at {\em two}-loops to the RGE's.}
\begin{eqnarray}
\frac{d}{dt} \left( \begin{array} {c}
m_{H_u}^2 (t) \\
m_{\tilde{t}}^2 (t) \\
m_{\tilde{t}^c}^2 (t)
\end{array} \right)
& = & \left( \begin{array} {c} 3 \\ 1 \\ 2 \end{array} \right)
\frac{\lambda_t^2 (t)}{8 \pi^2} \Bigl[ m_{H_u}^2 (t)
+ m_{\tilde{t}}^2 (t) + m_{\tilde{t}^c}^2 (t) + A _t ^2 (t) \Bigr]
+ \left( \begin{array} {c} B_{H_u} (t) \\ B_{\tilde{t}} (t) \\
B_{\tilde{t}^c} (t) \end{array} \right), 
\end{eqnarray}
where 
\begin{equation}
B_i (t) = - \frac{2}{\pi} \sum _A \alpha_A (t) C_A^i M_A^2 (t),
\end{equation}
$A_t$ is the trilinear soft SUSY breaking term,
$M_A$'s are the
gaugino masses, $C_A^i$ are Casimirs for the particle $i$
under the gauge group $A$ and
$t \sim \ln \mu_{RG}$  where $\mu _{RG}$ is the renormalization
scale.
The RGE for the top quark Yukwa coupling is
\begin{equation}
\frac{d \lambda _t(t)}{d t} = \frac{\lambda _t(t)}{16 \pi ^2} 
\Biggl[ 6 \lambda _t ^2 (t) + \sum_{A=1}^3 k_A g_A^2(t) \Biggr],
\end{equation}
where $(k_3, k_2, k_1) = ( -16/3, -3, -13/15)$ and $g_A$'s are the
gauge couplings.
The solution for $m_{H_u}^2$ can be written as
\begin{eqnarray}
m_{H_u}^2 (t) & = & m_{H_u}^2 (0) \left( \frac{1 + I}{2} \right) -
\left( \frac{1 - I}{2} \right)
\Bigl[ m_{\tilde{t}}^2 (0) + m_{\tilde{t}^c}^2 (0) \Bigr] \nonumber \\
 & & + a_{M_3} M_3^2 (0) + a_{M_2} M_2^2 (0) + a_{M_1} M_1^2 (0) \nonumber \\
 & & + a_{A_t} A_t^2 (0) + \sum_{A \neq B} a_{M_A \; M_B} M_A (0) M_B (0)
\label{mhu2sol}
\end{eqnarray}
with
\begin{eqnarray}
I & = & \hbox{exp} \left( \int _0 ^t dt^{\prime} 
\frac{6 \lambda ^2_t (t^{\prime}) }{8 \pi^2} \right) \nonumber \\
 & = & 1 + \frac{6}{8 \pi^2} \lambda _t ^2 (t) 
\frac{\int _0 ^t d t^{\prime} \prod_{A=1}^3 \left( 
\frac{\alpha _A (t^{\prime})}{\alpha _A(0)} \right) ^{(k_A/b_A)}}
{\prod_{A=1}^3
\left( \frac{\alpha _A (t)}{\alpha _A (0)} \right)^{(k_A/b_A)}},
\end{eqnarray}
$t = \ln
\left( \mu _{RG} / M_{mess} \right)$, where $M_{mess}$ is
the messenger scale (the RG scale at
which the SUSY breaking is mediated to the MSSM) and
$b_A$'s are the gauge beta-functions. 
Thus, $I$ (and the various $a$'s)
depends on $\lambda _t$ (or $\tan \beta$), the messenger
scale, $M_{mess}$ and the renormalization scale, $\mu _{RG}$.
Typically, $I \sim O(0.1 \; -1)$ and $|a_{M_3}| \sim O(1) >
|a_{M_{1,2}}|$.
Assume that the boundary condition is
\begin{equation}
m_{H_u}^2 (0) = y_{req} \Bigl[ m_{\tilde{t}}^2 (0) + m_{\tilde{t}^c}^2 
(0) \Bigr],
\label{bc}
\end{equation}
where
\begin{equation}
y_{req} \equiv \frac{1 - I}{1 + I}
\label{yreq}
\end{equation}
with $I$ evaluated at the weak scale, {\it i.e.},
$\mu _{RG} \sim O(100)$ GeV. Then,
we see from Eq. (\ref{mhu2sol}) that
$m_{H_u}^2 (\hbox{at} \;
\sim O(100 \; \hbox{GeV}))$ is independent of the 
boundary values of the scalar (Higgs and stop) masses
since there is a cancellation between the contributions of the
(bare) Higgs mass and stop masses -- we follow
the terminology of \cite{feng} and refer to this as
``focussing''.
Thus, as long as this boundary condition is satisfied
(in other words, there is focussing), the stop
masses can be large (multi-TeV)
{\em without} resulting in
$O(\hbox{TeV})^2$ $|m_{H_u}^2|$ at the weak scale.
Hence no large cancellation with $\mu^2$ is required to obtain the
observed $Z$ mass (see Eq. (\ref{mztree})), {\it i.e.},
the fine-tuning due to $\mu ^2$ {\em can} be small 
even though the stops are multi-TeV \footnote{ 
The other (gaugino mass) contributions to $m_{H_u}^2$ (at the
weak scale) (and also, if $\tan \beta$ is {\em small}, 
$m_{H_d}^2$ at the weak scale) have to
be $O (\hbox{a few} \; 100 \; \hbox{GeV})^2$ for $\mu^2$ to be $O(m_Z^2)$.}. 
Another fine-tuning in electroweak symmetry breaking
is due to the dependence of $m_{H_u}^2$ (at the weak scale)
and hence $m_Z$ (see Eq. (\ref{mztree}))
on the {\em stop mass}. 
For this fine-tuning to be small, it is
necessary, in general,
that the stop mass contribution to $m_{H_u}^2$ at the weak scale
be 
$O (100 \; \hbox{GeV})^2$. With multi-TeV stop masses, this contribution
is (typically) $O(\hbox{TeV})^2$ (see Eq. (\ref{mhu2sol}), where
$I \sim O(0.1 \; - \; 1)$). So, if the stop mass is varied keeping the
(bare) Higgs mass {\em fixed}, 
then $m_Z$ will be {\em very} sensitive to the (multi-TeV)
stop mass, even
though, due to a cancellation
between the contributions (both of which are $O(\hbox{TeV})^2$)
of the stop mass and bare Higgs mass,
we have
$|m_{H_u}^2|$ (at the weak scale) $\sim \mu^2 \sim O(m_Z^2)$. 
However,  
the sensitivity of
$m_{H_u}^2$ at the weak scale and hence $m_Z$ to variations of 
the (boundary value of) the stop mass (even if it is multi-TeV)
will be {\em small} as long as 
the Higgs mass is {\em also} varied according to
the boundary condition (Eq. (\ref{bc}))
so that the above mentioned cancellation
still takes place. 

Of course, the tree-level
relation of Eq. (\ref{mztree}) 
is modified by radiative corrections. In particular,
the one-loop contribution to the effective
Higgs potential depends on the stop masses.
The one-loop corrected expression for the $Z$ mass is
\begin{eqnarray}
\frac{1}{2} m_Z^2 & = & - \mu^2 + \frac{m^2_{H_d} + 2 \; \frac{\partial
\Delta V_1}{\partial v_d^2} - \left( m_{H_u}^2 + 2 \; \frac{\partial
\Delta V_1}{\partial v_u^2} \right) \tan ^2
\beta}{ \tan ^2 \beta -1} \nonumber \\
 & \approx & - \mu^2 - m_{H_u}^2  - 2 \; \frac{\partial
\Delta V_1}{\partial v_u^2} \; (\hbox{for large} \; \tan \beta),
\label{mzloop}
\end{eqnarray}
where $\Delta V_1$ is the one-loop contribution to the effective
potential. 
Keeping only
the stop and top mass contribution in $\Delta V_1$
and neglecting the mass mixing 
of the top squarks, we get
\begin{eqnarray}
2 \; \frac{\partial
\Delta V_1}{\partial v_u^2} & = & \frac{3 \; \lambda _t ^2}{32 \pi^2}
\Biggl[ -4 m_t^2 \Bigl[ \ln \left( \frac{m_t^2}
{\mu _{RG}^2}\right)  - 1 \Bigr] \Biggr. \nonumber \\
 & & \Biggl.+ 
2 \; \left( m_{\tilde{t}}^2 + m_t^2 \right)
\Bigl[ \ln \left( \frac{m_{\tilde{t}}^2 + m_t^2}
{\mu _{RG}^2} \right)  - 1 \Bigr] + 
2 \; \left( m_{\tilde{t}^c}^2 + m_t^2 \right)
\Bigl[ \ln \left( \frac{m_{\tilde{t}^c}^2 + m_t^2}
{\mu _{RG}^2}\right)  - 1 \Bigr]
\Biggr]. \nonumber \\
 & &
\label{oneloop}
\end{eqnarray}
Thus, 
even if the above boundary condition for the Higgs
and stop masses (Eq. (\ref{bc})) is satisfied, $m_Z$ will still depend
(weakly) on the stop masses.

It turns out that for the measured value of the top
quark (pole) mass, $m_t = 173.8 \pm 5.2$ GeV \cite{pdg}
and for $\tan \beta \stackrel{>}{\sim}
5$, $I \approx 1/3$ for RG scaling from the GUT or Planck scale to the
weak scale
\cite{feng}. This means that for the following boundary
condition in supergravity mediated SUSY 
breaking (where the messenger
scale is the GUT or Planck scale), ``focussing'' results:
\begin{equation}
m_{H_u}^2 (0) \approx \frac{1}{2} \Bigl[ m_{\tilde{t}}^2 (0) + 
m_{\tilde{t}^c}^2
(0) \Bigr].
\end{equation}
This class of models includes the minimal supergravity model where
there is a universal scalar mass ($m_0$) \cite{feng}. Hence, in minimal
supergravity and for these values of
$\tan \beta$ and $m_t$, scalars can 
be multi-TeV without leading to fine-tuning of $m_Z$.

\begin{figure}
\centerline{\epsfxsize=0.8\textwidth \epsfbox{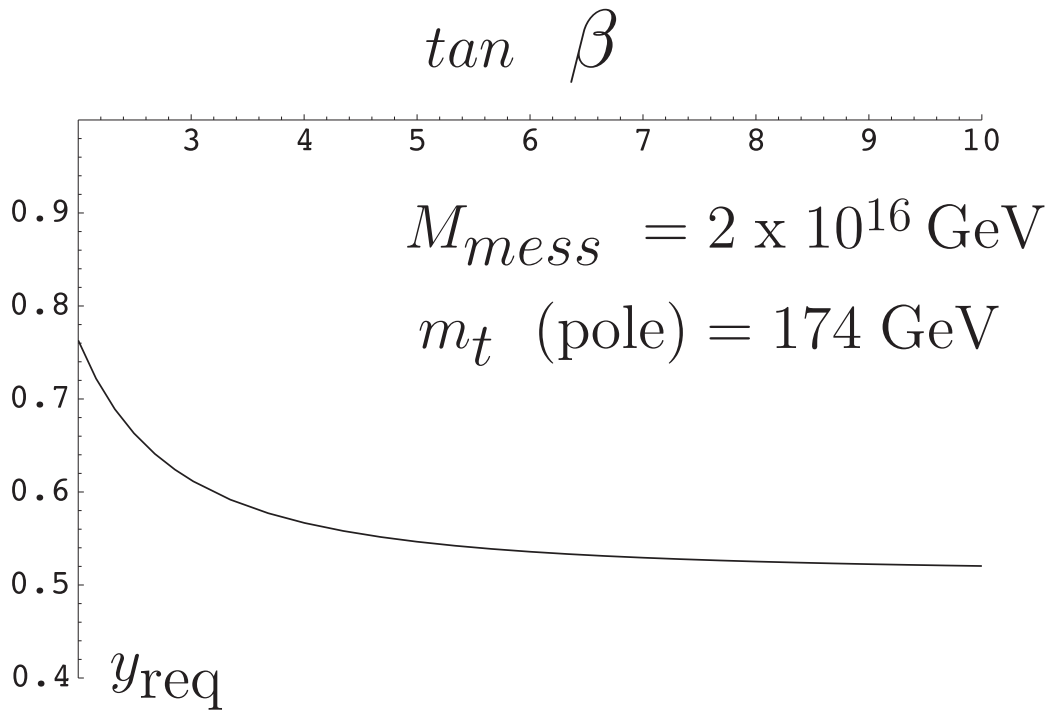}}
\caption{The ratio of the Higgs mass to the stop mass required
for focussing (see
Eq. (\protect\ref{yreq}))
as a function of $\tan \beta$
with the messenger scale $\sim 2 \times 10^{16}$ GeV.
The renormalization scale is
$\mu _{RG} \approx 1$ TeV and the top quark (pole) mass is
$174$ GeV.}
\protect\label{yreqsugra}
\end{figure}

It is clear from the expression for $I$ that for smaller $\tan \beta$
and hence larger $\lambda _t$,
$I$ will be smaller and hence $y_{req} > 1/2$
(for $M_{mess} \approx M_{GUT}$). Thus, even
for small $\tan \beta$
\footnote{Of course, there is a lower limit
on $\sin \beta \; (\tan \beta) $ of $\sim 0.85 \;
(1.6)$ if we impose the condition that 
$\lambda _t$ should not reach it's Landau pole below
the GUT scale.}, there is a boundary condition for the scalar masses
which leads to focussing 
in
supergravity mediated SUSY breaking so that,
even for small $\tan \beta$, stop masses can be multi-TeV without
resulting in fine-tuning.
The required value of $y$ is shown in 
Fig.\ref{yreqsugra} as a function of $\tan 
\beta$ where we see that $y_{req} \approx 1/2$ for $\tan \beta
\stackrel{>}{\sim} 5$.

\section{Minimal Gauge Mediation}
\label{minimal}
\begin{figure}
\centerline{\epsfxsize=1\textwidth \epsfbox{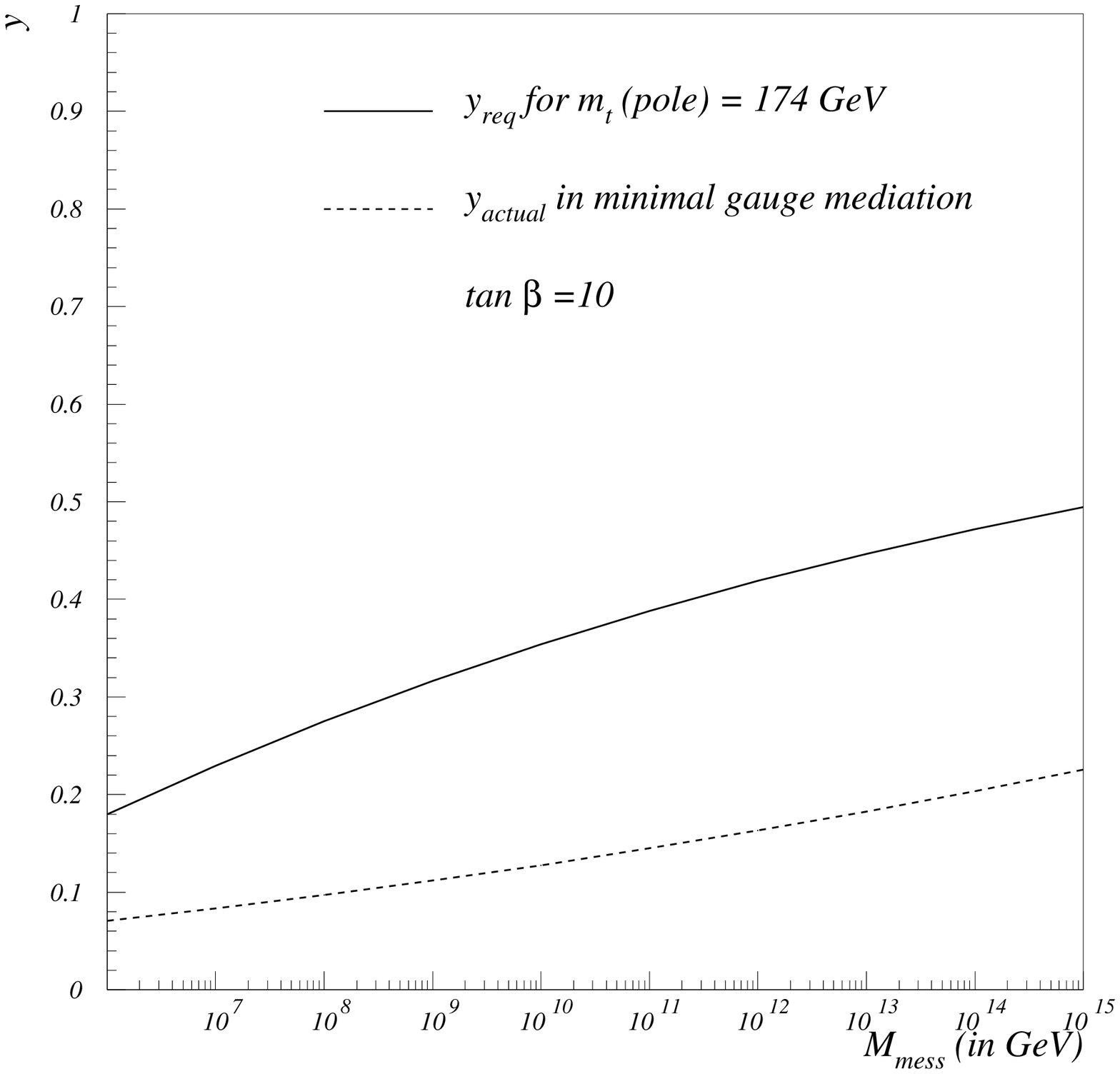}}
\caption{The
ratio of the Higgs mass to the stop mass {\em required}
for focussing (solid line)
(see
Eq. (\protect\ref{yreq})) 
as a function of the messenger scale, $M_{mess}$.
The renormalization scale is
$\mu _{RG} \approx 1$ TeV, $\tan \beta =10$ and the top quark (pole) mass is
$174$ GeV. The dashed line is 
the actual ratio of the Higgs mass to the stop mass
in the minimal model of 
gauge mediation (see Eq. (\protect\ref{yactual})).
}
\protect\label{ygmsb1}
\end{figure}

\begin{figure}
\centerline{\epsfxsize=0.8\textwidth \epsfbox{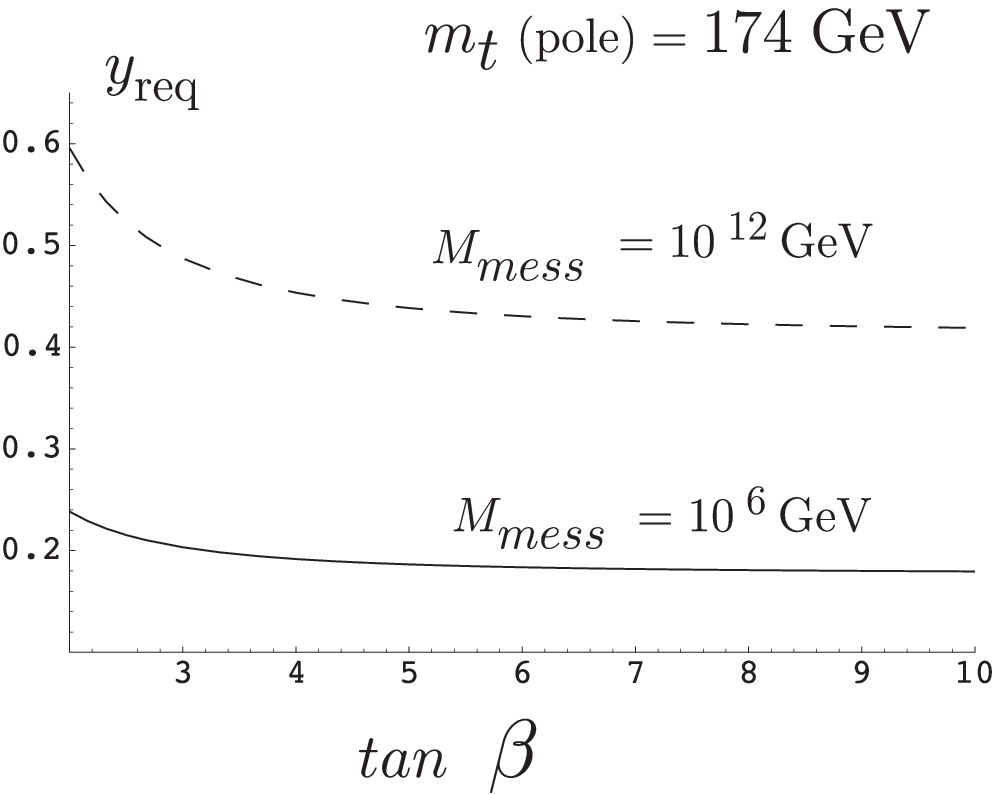}}
\caption{The 
ratio of the Higgs mass to the stop mass required
for focussing (see
Eq. (\protect\ref{yreq})) as a function of $\tan \beta$ for
messenger scale $M_{mess} = 10^6$ GeV (solid line) and
$M_{mess} = 10^{12}$ GeV (dashed line).
The renormalization scale is
$\mu _{RG} \approx 1$ TeV and the top quark (pole) mass is
$174$ GeV.}
\protect\label{ygmsb2}
\end{figure}

We now consider lower messenger scales, $M_{mess} \ll M_{Planck}$,
to see if ``focussing'' can take place. 
In Figs.\ref{ygmsb1} and
\ref{ygmsb2}, the required $y$ (Eq. (\ref{yreq})) is plotted as a function
of the messenger scale for $\tan \beta =10$ and as a function of 
$\tan \beta$ for a fixed messenger scale.

Gauge mediation is the only known
mechanism to mediate SUSY breaking to the MSSM at these energy scales. So,
we study the scalar mass predictions in gauge mediation to see if the 
boundary condition required for focussing can be satisfied.
Typically in models of gauge mediation, a SUSY breaking field $X$
(with non-zero
vev's in its scalar and $F$-components) 
couples to ``messenger''
fields which are vector-like under the SM gauge group.
When these
messenger fields are integrated out at the scale $M_{mess}
\sim X$, soft SUSY
breaking masses are generated for the MSSM scalars $(m_i^2)$
and gauginos
$(M_A)$:
\begin{eqnarray}
m_{\hbox{scalar} \; i}^2 & = & 2 \left( \frac{F_X}{X} \right)^2 \Biggl[
\sum _{mess,A} C_A^i N_{mess}^A \left( \frac{\alpha _A (X)}{4 \pi}
\right) ^2 \Biggr] \Biggl[ 1 + O \left( \frac{
F_X}{X^2} \right) \Biggr],
\label{mscalar}
\end{eqnarray}
\begin{eqnarray}
M_A & = & \frac{\alpha_A (X)}{4 \pi} \frac{F_X}{X} \sum _{mess} N_{mess}^A 
\Biggl[ 1 + O \left( \frac{
F_X}{X^2} \right) \Biggr],
\label{mgaugino}
\end{eqnarray}
where $N_{mess}^A \delta^{ab} = \hbox{tr} (T^a T^b)$, $T$'s are the
generators of the gauge group in the representation of the messengers
and $C_A^i$ is the Casimir of the scalar particle $i$
under the gauge group $A$.

To get an idea of focussing in a model, we define
\begin{eqnarray}
y_{actual} & \equiv & \frac{m_{H_u}^2 (0)}
{m^2_{\tilde{t}} (0) + m_{\tilde{t}^c} (0)}.
\label{yactualdef}
\end{eqnarray}
In a minimal model of gauge mediation,  a single SUSY breaking field
couples to 
messenger fields which are in complete multiplets of $SU(5)$
so that we get
\begin{eqnarray}
y_{actual} & = & \frac{ \frac{3}{4} \alpha ^2_2 (0) + \frac{3}{20}
\alpha _1^2 (0) }{ \frac{8}{3} \alpha ^2 _3 (0) + \frac{3}{4} \alpha ^2_2 (0)
+ \frac{17}{60} \alpha _1^2 (0) }.
\label{yactual}
\end{eqnarray}
In Fig.\ref{ygmsb1}, we also plot $y_{actual}$ (which depends
only on $M_{mess}$) for the minimal 
GM model: 
we see that for all $M_{mess}$, $y_{actual} < y_{req}$.
For smaller $\tan \beta$, $\lambda _t$ is larger and hence $y_{req}$
is larger. So, we see that for all $\tan \beta$
and $M_{mess}$, minimal GM does not
give the boundary condition required for ``focussing'' of $m_{H_u}^2$.

As mentioned in the introduction and as can be seen from Eqs. (\ref{mscalar}) 
and (\ref{mgaugino}), in typical models of GM
(in particular, the minimal model), the scalar and gaugino
masses are comparable so that multi-TeV scalars would imply
multi-TeV gauginos. In the minimal model,
$M_3 (0) > M_2 (0), \; M_1 (0)$ (for $M_{mess} \ll M_{GUT}$).
Thus, multi-TeV gauginos
(especially gluino)
will result in $m_{H_u}^2 \sim \; -\hbox{TeV}^2$
(see Eq. (\ref{mhu2sol}), where typically $a_{M_3} <0$
and for
$M_{mess}$ not too small,
$| a_{M_3} | \sim O(1)
> | a_{M_{1,2}} |$) and thus fine-tuning of $m_Z$
\footnote{As mentioned in the footnote on page 2, 
there has been extensive analysis
of EWSB in minimal GM from which it is clear that multi-TeV squarks
will lead to fine-tuning. We have redone (reworded) this analysis
so that the modifications to minimal GM required
to achieve ``focussing'' are transparent (see below).}.

Typically in GM (certainly in the minimal model), $m_{H_u}^2 \approx
m_{H_d}^2$ at $M_{mess}$ so that large ((multi-TeV)$^2$) $m_{H_u}^2$
at $M_{mess}$ 
implies that $m_{H_d}^2$ is also large.
For $\tan \beta \stackrel{>}{\sim}5$, $m_Z$ is very
{\em in}sensitive to $m_{H_d}^2$ (at the weak scale) (see Eq. (\ref{mztree}))
so that even if it is $\sim (\hbox{TeV})^2$, there is no fine-tuning,
{\it i.e.,} no large cancellation
with $\mu^2$ is required (as long as $m_{H_u}^2$ at the weak scale is not too 
large). However, if $\tan \beta$ is small and the Higgs masses
are $\sim$ TeV at $M_{mess}$, then at the weak scale, $m_{H_d}^2
\sim (\hbox{TeV})^2$ also
(since RG scaling results in a positive 
contribution to $m_{H_d}^2$
due to gaugino masses -- the Yukawa couplings of $H_d$ are small
for small $\tan \beta$); this results in
fine-tuning in EWSB (see Eq. (\ref{mztree}), where $m_{H_d}^2
\sim \hbox{TeV}^2$ and $\tan \beta$ is small).

Thus, in order to 
have multi-TeV squarks without resulting in
fine-tuning in models of GM,
we require:

{\bf 1.} additional contribution to the Higgs (mass)$^2$ (or
a non-minimal messenger sector) so that $y_{actual}$ (see Eq. 
(\ref{yactualdef}))
can be {\em larger} and thus focussing can be achieved. Also, either 
$\tan \beta \stackrel{>}{\sim}5$
so that a large ($\sim (\hbox{TeV})^2$)
$m_{H_d}^2$ (at the weak scale)
does not lead to fine-tuning or if $\tan \beta$ is small,
then the additional contribution
to the Higgs (mass)$^2$ should be such that
$m_{H_d}^2$ at the boundary is small ($O(100 \; \hbox{GeV})^2$)
even though
$m_{H_u}^2$ is larger 

{\em and}

{\bf 2.} 
{\em suppression} of the gaugino masses relative to the scalar masses 
so that
even if the scalars are multi-TeV, the gauginos can be sub-TeV 
and hence
not result in fine-tuning \\
or \\
a modification of the relation between the 
gaugino masses
so that the gaugino mass contribution to the Higgs (mass)$^2$ can be smaller
(say due to cancellation between the gluino and wino masses)
even if the gauginos are heavy.

\section{Non-Minimal Gauge Mediation}
\label{nonminimal}
\subsection{Increasing $m_{H_u}^2 (0)$}
\label{incrmhu20}
We discuss two ways to increase the Higgs mass relative to the stop
mass (at the messenger scale) in 
GMSB so that $y_{actual} \approx y_{req}$.

{\bf 1.} 
The Higgs potential in the MSSM has the parameters $B \mu$ and $\mu$
in addition to the Higgs soft (mass)$^2$. The parameter $B \mu$ breaks SUSY
and both $\mu$ and $B \mu$ break the $U(1)_{PQ}$ symmetry. In the
minimal model of GM, $B \mu$ is zero at the messenger scale and a non-zero
$B \mu$ (typically small) is induced in RG scaling. In generic models of GM,
there are additional interactions of the Higgs fields with the SUSY
breaking sector to (break
the $U(1)_{PQ}$ symmery and)
generate the $B \mu$ term (at $M_{mess}$) and also a $\mu$ term.
Typically the models which generate
the $\mu$ term have $\mu \approx
0$ in the supersymmetric limit so that the $\mu$
term is generated once SUSY is broken. Then, it's value is related
to the soft masses
and is therefore close to the weak scale as required by naturalness.

These additional interactions can also generate extra contributions
to the Higgs soft (mass)$^2$ ($m_{H_u}^2$ and $m_{H_d}^2$) \cite{higgsmass}.
In other words, there is effectively an extra parameter, $m_{H_u}^2 (0)$
in these models which is {\em independent} of the stop masses (which are
unchanged) \footnote{The ``effective'' $m_{H_u}^2 (0) =
m_{H_u}^2 (0)$ due to GM (this part is related to the stop
masses) +
$m_{H_u}^2 (0)$ due to additional Higgs interactions.}. 
If we choose this parameter so that $y_{actual} \approx y_{req}$
\footnote{As mentioned earlier, even if $y_{actual} = y_{req}$,
$m_Z$ will still be (weakly) sensitive to the stop mass due
to the one-loop contribution to the effective Higgs potential,
Eq. (\ref{oneloop}). We 
can define an ``effective'' $y_{req}$ to include this effect.}
then the Higgs (mass)$^2$ at the weak scale will be (almost)
independent of the stop masses, {\it i.e.,} it will
be $O(100 \; \hbox{GeV})^2$
even if the stop masses are multi-TeV. In other words,
for each $\tan \beta$ (or equivalently $\lambda_t$) and
$M_{mess}$ ({\it i.e.}, for each 
$y_{req}$) there is a value of $m_{H_u}^2 (0)$
which results in focussing 
\footnote{The additional Higgs interactions
generate $\mu$, $B \mu$ and a contribution to $m_{H_{u,d}}^2$
(in addition to the GM contribution).
Typically (see, for example,
the model of Dvali, Giudice, Pomarol in \cite{higgsmass})
$\mu$, $B \mu$ and (a part of) $m_{H_{u,d}}^2$ are given in terms of
the SUSY breaking scale (say, $m_{SUSY}$) and {\em more than
two} additional
Higgs couplings so that the two minimization conditions
for the Higgs potential (Eqs. (\ref{mzloop}) and (\ref{Bmu}))
are {\em not} sufficient to fix the values of these couplings
(for given $\tan \beta$, $M_{mess}$
and $m_{SUSY}$) and hence $m_{H_u}^2 (0)$
({\it i.e.}, $y_{actual}$) is a {\em free} parameter.}
\footnote{As mentioned
earlier, if $\tan \beta$ is small, then the additional Higgs
soft masses should be such that $m_{H_d}^2
\sim O(\hbox{a few} \; 100 \; \hbox{GeV})^2$ at $M_{mess}$.}. 

Of course, this requires a ``coincidence'', {\it i.e.,}
a correlation between the additional Higgs interactions
which generate $m_{H_u}^2 (0)$ (the GM contribution
to $m_{H_u}^2 (0)$ relative to the stop masses is fixed for
a given messenger sector)
and the messenger scale and/or $\lambda_t$ (since
$y_{req}$ depends on the messenger scale and $\lambda_t$).
This means that, with
(boundary) scalar masses $\sim$ multi-TeV, if these 
additional Higgs couplings (and hence only $m_{H_u}^2 (0)$)
are changed by $O(1)$, then the Higgs (mass)$^2$
at the weak scale changes by $O(\hbox{TeV})^2$ so that $m_Z$
is very sensitive to changes in these couplings.
On the other hand, if $m_{SUSY} \sim \alpha / (4 \pi) \;
F_X / X$ which determines
the size of the soft masses for the MSSM sparticles is changed by
$O(1)$ (with the additional Higgs couplings fixed), then both the Higgs
{\em and} the stop masses change such that $y_{actual}$ remains the
same and thus $m_Z$ is {\em in}sensitive to changes in $m_{SUSY}$ even
if $m_{SUSY}$ is multi-TeV
(of course, provided $y_{actual} \approx y_{req}$).

This situation is similar to the case of focussing in supergravity
mediated SUSY breaking with universal scalar masses (for $\tan \beta
\stackrel{>}{\sim}5$) \cite{feng}. 
As long as the Higgs {\em and} the stop masses
are varied keeping the boundary
relation between the masses the same as the one which
results in focussing
(for example,
by changing the scale of SUSY breaking), 
$m_Z$ is not very sensitive to these
masses, even if they are multi-TeV. 
But, there is, a priori (in the absence of
flavor or other symmetries), no reason for
scalar masses to be universal (or to satisfy the
particular boundary condition)
in supergravity
so that in generic 
supergravity models it should be possible to vary the stop mass
keeping the (bare) Higgs mass {\em fixed}. As discussed earlier,
in this case, if the stop mass is multi-TeV,
then $m_Z$ {\em is} very sensitive to it (even though 
the Higgs (mass)$^2$ at the weak scale 
may be $O(100 \; \hbox{GeV})^2$).

{\bf 2.} If the number of messenger fields charged under $SU(2)_{weak}/
U(1)_Y$ (messenger ``leptons'')
is larger than those charged under $SU(3)_{color}$ (messenger
``quarks''),  
then the Higgs mass (relative to the stop masses) will be
larger (see Eq. (\ref{mscalar})). 
Other possibilities to increase $y_{actual}$ are to
have a smaller supersymmetric mass 
($X$ in Eq. (\ref{mscalar})) 
for the messenger leptons (see section \ref{raby} for
a specific model)
or to have a larger SUSY breaking mass 
($F_X$ in Eq. (\ref{mscalar})) 
for the messenger leptons (see section \ref{dtwmodel} for a specific
model).
In the case where
the numbers (or the supersymmetric
masses) of the messenger quarks and leptons are not equal, 
unification of the SM gauge couplings
will be affected.
Unlike the case 1 above, the particular boundary relation between
the Higgs and the stop masses is guaranteed by the content
of the messenger sector, {\it i.e.,} once the messenger sector is 
fixed, we (typically)
cannot vary the Higgs and stop masses {\em independently}.

\subsection{Reducing the fine-tuning due to gaugino masses}
In this section, we discuss some methods to reduce the gaugino mass
contribution to the Higgs (mass)$^2$ at the weak scale in models of GM
so that even if the 
scalars are multi-TeV, the gaugino masses will not lead to fine-tuning
in EWSB.

{\bf 1.} It is possible that the gaugino masses are generated by a 
set of messenger quarks and leptons which are different from
those that generate the scalar masses. This means that
there can be an 
extra (independent of the scalar masses) parameter for gaugino masses. For 
example, in the model of Dobrescu \cite{dob}, there is one
set of messengers which are SM singlets, but charged under $U(1)_{(B-L)}$ 
gauge group. SUSY breaking is mediated
to the MSSM squarks and sleptons, but {\em not} to the MSSM gauginos
(and the Higgs fields), 
by $(B-L)$ gauge interactions. There is another set of messengers which
are vector-like under the SM gauge group. This second set of messengers
gives 
contributions to {\em both} scalar and gaugino masses
which depend on different (superpotential) couplings (and also might
be smaller by a loop factor) than
the $U(1)_{(B-L)}$ contribution (to scalar masses)
\footnote{In \cite{mohapatra} also,
there is $U(1)_{(B-L)}$ gauge mediation (which
generates SUSY breaking masses for squarks and sleptons only)
in addition to mediation by SM gauge
group (which is necessary to
generate MSSM gaugino masses), 
but both these contributions are related and hence
gaugino masses are {\em not} independent of scalar masses.}.
Thus, 
effectively, there are two parameters, one for the gaugino masses and one for 
the scalar masses so that even if scalars are multi-TeV, the gauginos can be
light.

{\bf 2.} The gaugino masses break $U(1)_R$ symmetry so that generically
the gaugino mass generated at one-loop by
integrating out messenger quarks and leptons
is given by $\sim \alpha / (4 \pi) F 
M_{\not \! R} / M_{mess}^2$, where $F$, 
$M_{\not \! R}$ and 
$M_{mess}$ are, respectively, the SUSY breaking, $R$-symmetry
breaking and messenger scales. In minimal models of gauge mediation,
$M_{\not
\! R} \sim M_{mess}$ so that there is no suppression of the gaugino masses
relative to the scalar masses (which do not break the $R$-symmetry). 
But, in non-minimal models, it is possible that $M_{\not
\! R} \ll M_{mess}$ so that (even if the {\em same} 
messengers generate scalar and 
gaugino masses) the gaugino masses are suppressed
by the ratio of
$R$-symmetry breaking scale to the messenger scale.
Thus, again, there is effectively
an extra (independent) paramater for gaugino masses. An example of this
idea is shown in section \ref{dtwmodel} as part of a specific model.

{\bf 3.} As in case 2 of section \ref{incrmhu20}, if the number of
messenger leptons is more than the number of quarks or
their masses (supersymmetric or SUSY breaking) are different, then
the wino mass relative to gluino mass can increase
(see Eq. (\ref{mgaugino})).
In this case, the net gaugino mass
contribution to the Higgs (mass)$^2$ is not too
large in magnitude ({\it i.e.}, it can be $O(\hbox{a few} \; 100 \;
\hbox{GeV})^2$)
even if the gauginos are heavy (say $\sim$ TeV)
(see Eq. (\ref{mhu2sol}),
where $a_{M_3} < 0$,
$a_{M_2} >0$ and $|a _{M_3}| >
|a_{M_2}|$). In other words, if the ratio of wino mass to gluino mass
is larger, then 
there can be a (partial) cancellation between
wino and gluino mass 
contributions to $m_{H_u}^2$ at the weak scale. A concrete 
model with $3$ pairs of messenger leptons and $1$ pair of quarks is
analysed in section \ref{qlmodel}.

{\bf 4.} For very low messenger scales, $|a_{M_A}|$'s in Eq. (\ref{mhu2sol})
are small so that gauginos
can be heavy without resulting in large $|m_{H_u}^2|$ at the weak scale
and hence fine-tuning.

For electroweak symmetry breaking to occur, (typically)
we require $m_{H_u}^2$
(at the weak scale) to be negative: 
either the scalar mass or gaugino
mass contribution to $m_{H_u}^2$ at the weak scale has to be negative. 
Thus, too much ``focussing'' 
({\it i.e.,} too small scalar mass contribution)
{\em and}
too much suppression of the gaugino masses (especially gluino relative
to wino) will not result in EWSB. For multi-TeV squarks to be
natural, 
it is {\em not} necessary that the
scalar mass contribution to $m_{H_u}^2$ at the weak scale vanish exactly:
EWSB is natural as long as both the scalar and gaugino mass contributions
are $O(\hbox{a few} \; 100 \; \hbox{GeV})^2$ 
and {\em at least} one of them is
negative so that $m_{H_u}^2 \sim - O(100 \; \hbox{GeV})^2$. It is also
possible that both the contributions are $O(\hbox{TeV})^2$ (again, with
at least one of them negative), 
but there is
a cancellation so that $m_{H_u}^2 \sim - O(\hbox{a few} \; 
100 \; \hbox{GeV})^2$.

\section{Specific Models}
\label{specific}
We now study concrete models of gauge mediation
which incorporate the ideas of section 
\ref{nonminimal}
and thus have multi-TeV squarks without leading to fine-tuning in EWSB.
\subsection{Raby model}
\label{raby}
In this model \cite{raby, rabytobe}, the messenger quarks have a 
(supersymmetric) mass $\sim M_{GUT}$
\footnote{This
model implements the Dimopoulos-Wilczek doublet-triplet splitting
mechanism \cite{dw}: 
a pair of messenger quarks has a Dirac mass with the 
pair of Higgs triplets due to
the vev of the $SO(10)$ adjoint field while the Higgs doublets
are massless.} whereas (some of) the messenger leptons have a smaller
(supersymmetric)
mass, $M \sim 10^{15}$ GeV. The $R$-symmetry breaking scale, $M_{\not \!R}$,
is $\sim M$. Thus, the gluino mass is suppressed relative to the wino mass
(and left-handed slepton mass)
by 
$\left( M / M_{GUT} \right)^2$: one factor of $M / M_{GUT}$ is due to 
the larger messenger scale ($M_{mess} \sim 
M_{GUT}$) for quarks and the other is due to 
$M_{\not \!R}$ being smaller than $M_{mess}$ (whereas for wino mass,
$M_{mess} \sim M_{\not \!R} \sim M$). As mentioned earlier, this means that 
even if the
left-handed slepton/squark masses are multi-TeV, the gluino can be lighter.
Also, the right-handed stop (and other squark) (mass)$^2$ is
suppressed by $\left( M / M_{GUT} \right) ^2$ relative to the left-handed
squark/slepton (mass)$^2$ due to the larger supersymmetric mass for messenger
quarks. This increases $y_{actual}$ as compared to minimal models of GM.
Thus, this model has the features required to have multi-TeV squarks
without resulting in fine-tuning. 

However, as shown in \cite{rabytobe}, it turns out that the suppression of
the gluino mass and the right-handed stop mass is too much if
$M \sim 10^{15}$ GeV or smaller:
one problem is that
both the gaugino and the scalar mass contributions to the 
(up-type) Higgs (mass)$^2$
at the weak scale tend to be positive
(wino mass is large relative to gluino mass
and $y_{actual}$ is {\em too} large) so that it is
hard to get EWSB (in particular, $\mu ^2$ tends to be too
small or negative unless
$\tan \beta$ is small while for small $\tan \beta$, the lightest
Higgs mass is too small).
The other problem is that
right-handed stop (mass)$^2$ is driven negative by the top quark
Yukawa coupling
in RG scaling to the weak scale
since it's boundary value is suppressed compared to
the Higgs and left-handed stop mass and also the light gluino
does not give a large enough positive contribution.
If the supergravity contribution to the soft masses (which is
$O(F_X / M_{Planck} )$, but with, in general, arbitrary coefficients)
is to be smaller than the
GM contribution, we need $M \stackrel{<}{\sim} 1/10 M_{GUT} \sim
10^{15}$ GeV;
otherwise, the soft mass
spectrum is not predictive.
Thus, it is hard to make this model phenomenologically viable in the desired
region of parameter space. 

In \cite{rabytobe}, these problems were alleviated by adding a $U(1)$
$D$-term contribution to the soft
scalar masses -- the additional contribution
is negative (positive) for the Higgs (squark) soft (mass)$^2$. However,
this {\em reduces} $y_{actual}$ and thus ``focussing'' is hard to
achieve.
 
\subsection{More messenger leptons than quarks}
\label{qlmodel}
As mentioned earlier
(see Eqs. (\ref{mscalar}) and (\ref{mgaugino})), 
if the number of messenger leptons is larger than
messenger quarks, then (a) $y_{actual}$ is larger so that the magnitude of
the scalar mass
contribution to $m_{H_u}^2$ at the weak scale decreases and (b)
the wino mass (relative to the gluino mass) increases, thus decreasing the 
(magnitude) of the gaugino mass 
contibution to $m_{H_u}^2$ at the weak scale
as well. Then, the squarks and some of the gauginos can be heavier than 
$1$ TeV without leading to fine-tuning in EWSB. A model of GM
with $3$ pairs of messenger leptons, denoted by $l + \bar{l}$, 
and $1$ pair of messenger quarks, denoted by
$q + \bar{q}$, where $q + l$ form a ${\bf 5}$ of
$SU(5)$,
and with a low 
messenger scale $\sim 100$ TeV was used in \cite{graesser1} to reduce the 
fine-tuning in EWSB. Here, we consider the same model,
but with higher messenger scales also. The soft mass spectrum
at the messenger scale is (assuming $\sqrt{F_X} \ll X$):
\begin{eqnarray}
m_{\hbox{scalar}\; i} & \approx & 2 \; \left( \frac{F_X}{X} \right)^2 \Biggl[ 
C^i_3 \left( \frac{\alpha _3}{4 \pi} \right) ^2 +
3 \; C^i_2 \left( \frac{\alpha _2}{4 \pi} \right) ^2 + \frac{3}{5}
Y_i^2 \; \frac{11}{5} \left( \frac{\alpha _1}{4 \pi} \right) ^2 \Biggr],
\end{eqnarray}
\begin{eqnarray}
M_3 & \approx & \frac{\alpha _3}{4 \pi} \frac{F_X}{X}, \nonumber \\
M_2 & \approx & 3 \; \frac{\alpha _2}{4 \pi} \frac{F_X}{X}, \nonumber \\
M_1 & \approx & \frac{11}{5} \; \frac{\alpha _1}{4 \pi} \frac{F_X}{X},
\label{mgaugino31}
\end{eqnarray}
where the gauge couplings are evaluated at $M_{mess} \sim X$.

With the above messenger sector ({\it i.e.},
$3$ pairs of leptons and $1$ pair of quarks)
and the MSSM particle content, the
three gauge couplings do {\em not} unify at $\sim 10^{16}$ GeV. However,
in \cite{graesser1}, 
with a messenger scale $\sim 100$ TeV,
color triplets were added with a mass of $\sim$ TeV
so that coupling unification can be maintained \footnote{
These exotic ``quarks'' were used to generate $\mu$ and 
$B \mu$ terms with the addition of a gauge singlet. Also, this
particle content could be embedded in a GUT model with a doublet-triplet
splitting mechanism \cite{graesser1}.}. Here, we do {\em not}
introduce any additional quarks to preserve coupling unification.
 
In this model, all the gaugino and the scalar masses can be written in 
terms of only one parameter 
-- we choose this parameter to be the gluino mass at the 
messenger scale, $M_3 (0)$. Thus, $y_{actual}$ is 
{\em not} useful to determine if
we can have focussing, {\it i.e.,} {\em both} the scalar mass and gaugino
mass contibutions to $m_{H_u}^2$ at the weak scale depend on
$M_3 (0)$.
We assume that there is a mechanism to generate 
$\mu$ and $B \mu$ terms and that this mechanism does {\em not}
give additional contributions to the Higgs soft masses. 
We also assume that no trilinear $A$-term is
generated at the messenger scale (as in the minimal model of GM).
Thus, the
{\em fundamental} parameters of this model are $M_3 (0)$, $M_{mess}$,
$\lambda _t$, $\mu$ and $B \mu$: these parameters are constrained by
the measured values of $m_Z$ and $m_t$ so that there are
$3$ {\em free} (or input)
parameters which we choose to be $M_3 (0)$, $M_{mess}$
and $\tan \beta$.
The values of $\mu$ and $B \mu$ are determined 
(in terms of $\tan \beta$ and
$M_3(0)$) from the minimization
conditions, Eqs. (\ref{mzloop}) and (\ref{Bmu}) (see below),
for the Higgs potential. 
\begin{equation}
\sin 2 \beta = \frac{2 B \mu}{ 2 \mu^2 + m_{H_u}^2 + m_{H_d}^2 }.
\label{Bmu}
\end{equation}

For the analysis
in this and the next section,
we use {\em one-loop} RGE's to evolve the scalar and gaugino masses and
the dimensionless couplings
from $M_{mess}$ to the weak scale. 
The one-loop contribution to the effective
Higgs potential (Eq. (\ref{oneloop})) is taken into account.
If the scalar soft masses
are $1 \; - \; 3$ TeV at $M_{mess}$, then, in RG scaling to the weak scale,
the {\em two-loop} contribution (due to the heavy scalar masses)
to the Higgs
and the squark/stop (mass)$^2$ (depending on
$M_{mess}$) can be $O (100 \; \hbox{GeV}
)^2$ (it is
smaller for the Higgs (mass)$^2$ than the stop (mass)$^2$)
\footnote{For numerical estimates of this two-loop
contribution, see, for example, \cite{graesser2}.}.
This contribution is typically {\em positive} for the Higgs (mass)$^2$
(due to the effect of
$\lambda _t$) and thus improves the fine-tuning, {\it i.e.}, reduces
$c$ defined below by $O(1)$ and so does not change the
order-of-magnitude estimate of fine-tuning.
The two-loop contribution to the 
the squark/stop (mass)$^2$ is {\em negative}, but since the 
squark/stop (mass)$^2$
after one-loop RG evolution to the weak scale is $\sim
O(\hbox{TeV})^2$ (for multi-TeV boundary scalar masses), the
two-loop
effect is not significant.

\begin{figure}
\centerline{\epsfxsize=1\textwidth \epsfbox{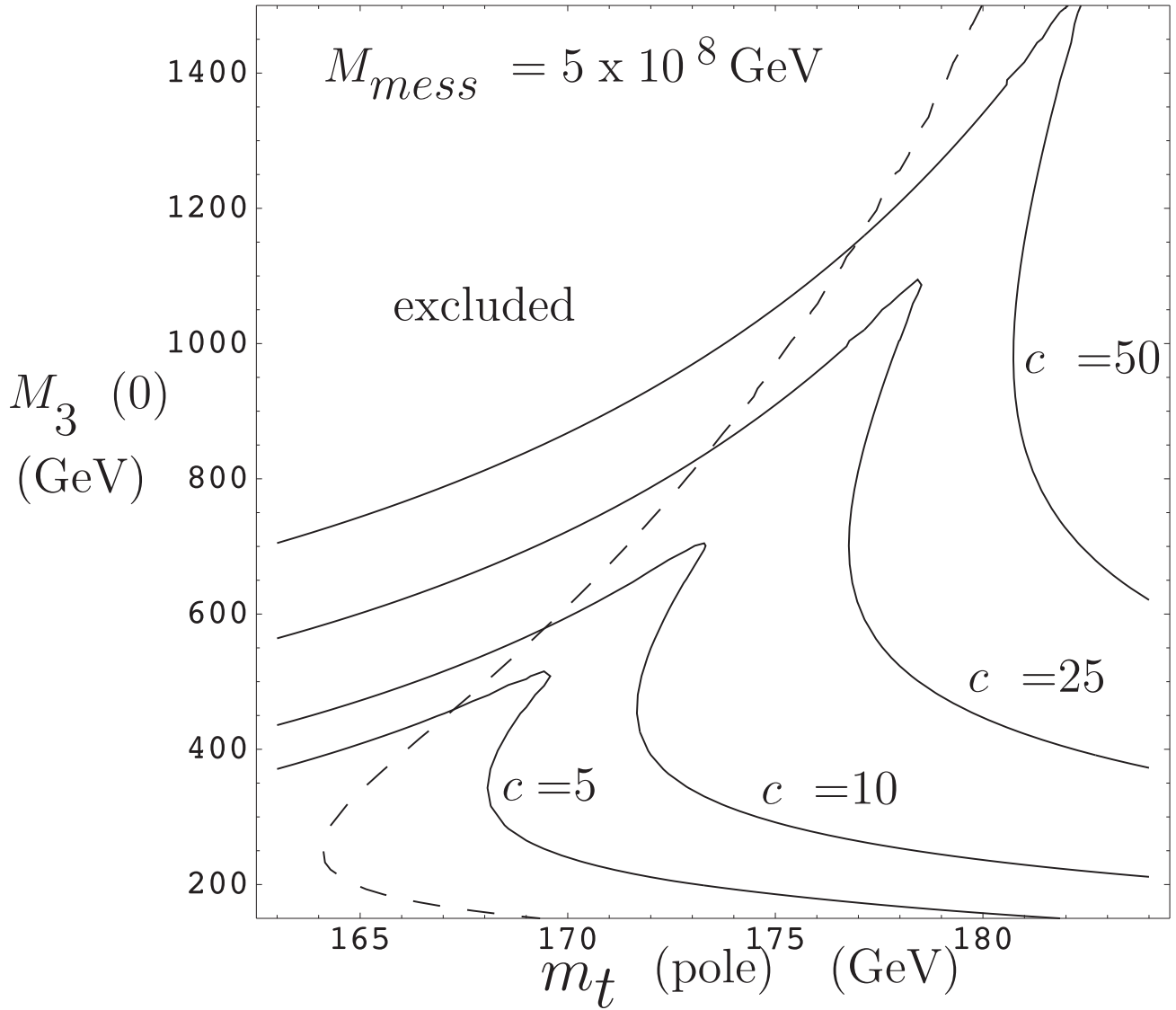}}
\caption{The fine-tuning $c$ (Eq. (\protect\ref{c})) (solid lines)
in the model of GM with $3$ pairs of
messenger leptons and $1$ pair of messenger quarks 
(section \ref{qlmodel}) as a function of
the gluino mass at the messenger scale, $M_3 (0)$, and the top quark 
(pole) mass, for 
$\tan \beta = 10$. 
The messenger scale is $\approx 5 \times 10^8$ GeV.
The region to the left of the dashed line has chargino mass
smaller than $90$ GeV or $\mu^2 < 0$ and is therefore excluded.}
\protect\label{qlM30mt}
\end{figure}

We use the sensitivity of $m_Z$ to a fundamental parameter,
$\lambda _i$, as a measure of fine-tuning due to that parameter
\cite{bg}:
\begin{equation}
c (m_Z^2; \lambda _i) \equiv \frac{\lambda _i}{m_Z^2}
\frac{\partial m_Z^2}{\partial \lambda _i}.
\end{equation}
It has been argued that a large value of $c(m_Z^2; \lambda _i)$ 
does not necessarily
imply that $m_Z$ is fine-tuned \cite{ac}. However, in this paper,
we are interested in order-of-magnitude estimates of fine-tuning only
and so we use 
\begin{equation}
c
\equiv \hbox{Max} \Bigl[ | c(m_Z^2;
\mu^2)|, |c(m_Z^2; m^2_{SUSY})| \Bigr],
\label{c}
\end{equation}
where $m_{SUSY}$ is a SUSY breaking
parameter in the MSSM,
as a measure of fine-tuning: $c \stackrel{>}{\sim}
100$ signals unnaturalness. 
If the stops are multi-TeV, then
the fine-tuning due to the top quark Yukawa coupling  
($c(m_Z^2; \lambda_t$)) (with
$\lambda_t$ evaluated at $M_{mess}$) is large
since
the stop mass contribution to $m_{H_u}^2$
at the weak scale (and hence to $m_Z^2$) is $O(\hbox{TeV}^2)$ (even
if $m_{H_u}^2$ (at the weak scale) $\sim m_Z^2$) \footnote{
To be precise, $m_{H_u}^2$ (at the weak scale) $\sim
-O(\hbox{a few} \; 100 \; \hbox{GeV})^2$ due to a cancellation between
$\sim O(\hbox{TeV}^2)$ contributions of the Higgs and stop
masses. Thus, if $\lambda _t$ is varied by (say) a few percent, then
this
cancellation is no longer {\em as} effective so that
$m_{H_u}^2$ (and hence $m_Z^2$) changes by $O(1)$.
This can be seen from Figs.\ref{qlM30mt} and \ref{dtwmucmt} where
we see that $c \sim \mu^2 / m_Z^2 \sim m_{H_u}^2 / m_Z^2$
is very sensitive to $m_t$ (for a fixed
$\tan \beta$) or in other words $m_{H_u}^2$ at the weak scale
is very sensitive to $\lambda _t$.}.  
We choose {\em not} to include this fine-tuning in $c$
as in \cite{feng}:
as argued by those authors, it is possible that the top
quark Yukawa coupling reaches a ``fixed point'' in RG scaling from
say the Planck scale to the GUT scale or $M_{mess}$ so that
the sensitivity to $\lambda _t (M_{Pl})$ (which is the fundamental
parameter) may not be large. Also $\lambda _t$
is {\em not} related to SUSY breaking -- we are interested mainly
in the fine-tuning due to the SUSY breaking parameters (we include
$\mu$ in this since in many models, $\mu$ {\em is}
related to SUSY breaking).
 
\begin{figure}
\centerline{\epsfxsize=1\textwidth \epsfbox{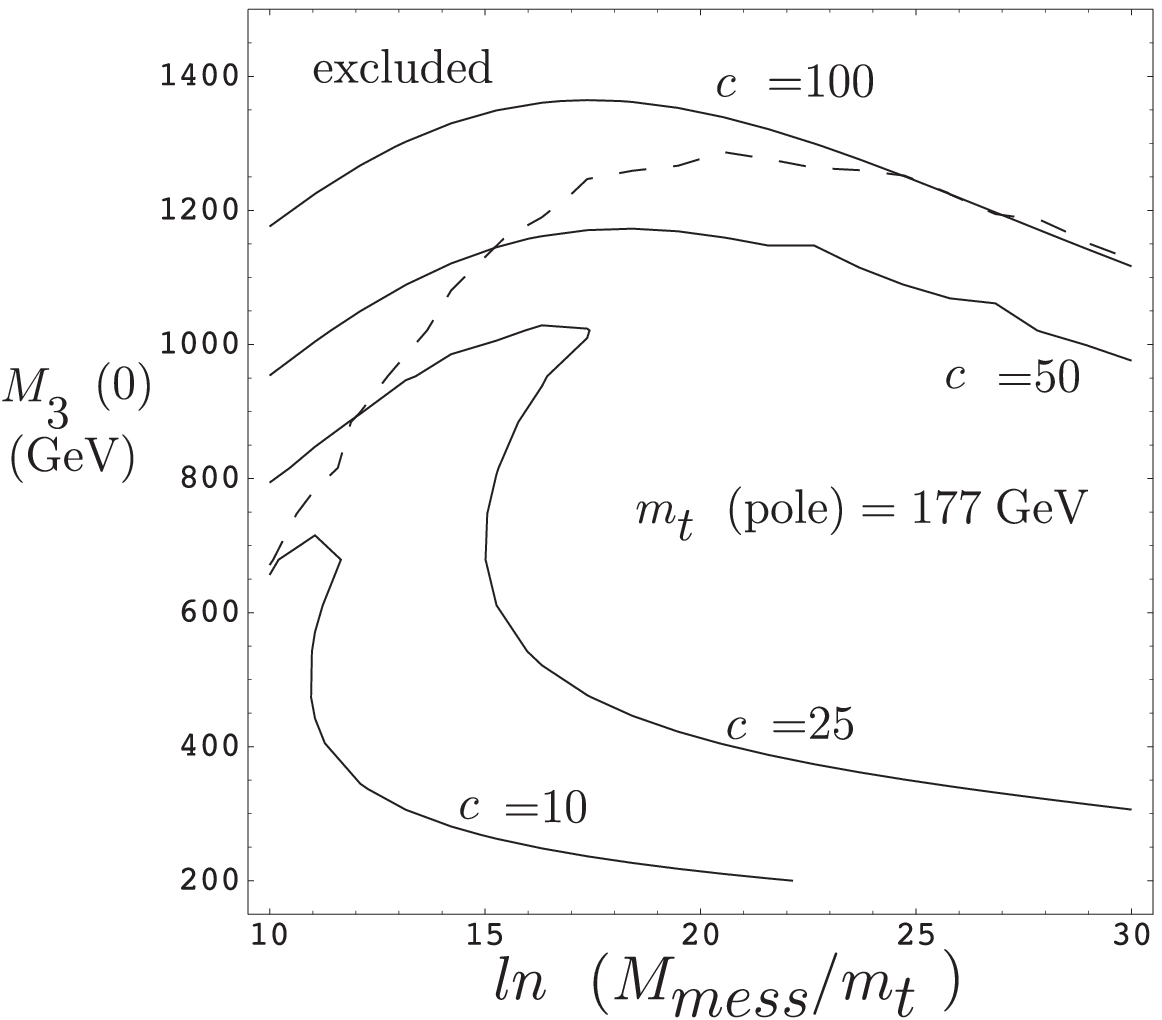}}
\caption{The fine-tuning $c$ (Eq. (\protect\ref{c})) (solid lines)
in the model of GM with $3$ pairs of
messenger leptons and $1$ pair of messenger quarks 
(section \ref{qlmodel}) as a function of
the gluino mass at the messenger scale, $M_3 (0)$, and the messenger scale, 
$M_{mess}$, 
for $m_t$ (pole) $= 177$ GeV and $\tan \beta = 10$. 
$\ln (M_{mess} / m_t) = 10$ corresponds to $M_{mess} \approx
4 \times 10^6$ GeV and $\ln (M_{mess} / m_t) = 30$ corresponds to
$M_{mess} \approx 2 \times 10^{15}$ GeV. 
The region above
the dashed line has chargino mass
smaller than $90$ GeV or $\mu^2 < 0$ and is therefore excluded.}
\protect\label{qlt0M30}
\end{figure}

If Figs.\ref{qlM30mt} and \ref{qlt0M30}, 
we show contours of fine-tuning in this model. 
The renormalization scale, $\mu _{RG}$, is chosen to be
$2 \; M_3(0) \approx \sqrt{m_{\tilde{t}_1} m_{\tilde{t}_2}}$ (where
$m_{\tilde{t}_{1,2}}$ are the physical stop masses) \footnote{
With the one-loop contribution to the Higgs potential included,
the results will not be very sensitive to the RG scale.}.

The right-handed squark (excluding the stop)
masses (at the weak scale) are $\approx 2 M_3(0)$
whereas the left-handed squarks (including the stop) are heavier,
where $M_3(0)$ is the gluino mass at the messenger scale.
The right-handed stop (to be precise,
the lighter stop eigenstate)
mass is (depending on $M_{mess}$)
$(1 \; - \; 2) \times M_3(0)$,
{\it i.e.}, it is smaller than the other squarks
(due to RG scaling to the weak scale with a large Higgs mass at the
boundary).
The right-handed slepton mass is (depending on $M_{mess}$) 
$(1/2 \; - \; 3/2) \times M_3(0)$. Gaugino masses at the weak scale can be 
obtained (in terms of $M_3(0)$)
from Eq. (\ref{mgaugino31}) 
using $M_A (t) \propto \alpha _A (t)$
(at one-loop).

From these figures, 
we see that even for $M_3 (0) \stackrel{>}{\sim} 500$ GeV,
it is possible to have $c \stackrel{<}{\sim} 100$ for a wide range of
messenger scales, especially for 
values of $m_t$ (slightly) {\em larger} than the central value  
(but within
$\sim 1 \sigma$ of the central value). 
Also, for some values of the other parameters,
the fine-tuning for 
$1.5$ TeV $
\stackrel{>}{\sim}
M_3 (0) \stackrel{>}{\sim} 500$ GeV can be almost as good as for
$200$ GeV $ \stackrel{<}{\sim} M_3 (0) \stackrel{<}{\sim} 500$ GeV 
(the lower limit on $M_3 (0)$ is from collider searches
for right-handed sleptons and stops)
\footnote{For example, for values of $m_t$ about $1 \sigma$ larger than
the central value, we see in Fig.\ref{qlM30mt} that the fine-tuning
contours are sort of parallel to the $M_3 (0)$ axis.}.
Thus, multi-TeV ($\sim
1 \; - \; 3$ TeV) squarks (including the left-handed stop) can be natural
in this model: discovery of these
sparticles at the LHC might be difficult
(at least with $10$ (fb)$^{-1}$ of integrated luminosity).
For
$M_3 (0) \stackrel{<}{\sim} 1$ TeV 
the right-handed stop mass can be
$\stackrel{<}{\sim} 1$ TeV so that it is likely to
be detected at the LHC. 
For $M_3 (0) \stackrel{>}{\sim} 500$ GeV, (depending on $M_{mess}$)
the gluino mass (at the weak scale)
can be $\stackrel{>}{\sim} 1$ TeV 
so that it might not
be detected at the LHC with $10$ (fb)$^{-1}$ of integrated luminosity. 
The wino mass (at the weak scale)
is about the same as the gluino mass, but for
the region of parameter space with $c \stackrel{<}{\sim} 100$,
the $\mu$ term is $\stackrel{<}{\sim} 600$ GeV so that even if the wino
mass is $\sim 1$ TeV
or more, 
the lighter chargino/neutralino will still be detected at a ($1$ or $1.5$)
TeV 
lepton collider (or at the LHC, if $\mu$ is very small), 
even though the heavier chargino/neutralino
will escape detection.
The right-handed slepton masses, even
for $M_3 (0) \stackrel{>}{\sim} 500$ GeV
(which leads to squarks heavier than 
$\sim 1$ TeV), can be lighter than $\sim
500 \; - \; 750$ GeV so that their discovery is possible at a 
$\sqrt{s} = 1$ or $1.5$ TeV lepton collider
\footnote{As mentioned earlier, right-handed sleptons
are lighter than squarks 
(by a factor
$\sim
\alpha _1
/ \alpha _3$ with $\alpha$'s evaluated at $M_{mess}$) as in typical
models of GM.}. 
The lightest SM superpartner
is typically the neutralino which has
bino as the dominant component
since $M_1 \sim 0.4 M_3$ (at the weak scale).

The results presented for $\tan \beta =10$ are also valid 
(roughly) for 
$5 \stackrel{<}{\sim} \tan \beta \stackrel{<}{\sim}20$
since in this range of $\tan \beta$, $\lambda _t$ does not change much
while the bottom and tau Yukawa couplings are small.
For smaller values of $\tan \beta$
and $M_3 (0) \stackrel{>}{\sim}500$
GeV, the fine-tuning will be worse than for large $\tan \beta$
since then $m_{H_d}^2 \sim$ (TeV)$^2$ (in this model, the two 
Higgs masses are the same at $M_{mess}$).

For very low messenger scales 
(and with multi-TeV squarks) so that the RG logarithm
is small and for $N_l \gg N_q$, it is hard to get
$m_{H_u}^2$ (at the weak scale) $<0$ since the positive wino mass and (bare)
Higgs mass contributions
to $m_{H_u}^2$ at the weak scale
dominate. This can be seen in Fig.\ref{qlt0M30}
where the region with very small $M_{mess}$ 
($\sim 10^6$ GeV) and $M_3 (0) \stackrel{>}{\sim}
500$ GeV is excluded since it has $\mu ^2 < 0$.

\subsection{Dimopoulos, Thomas, Wells model}
\label{dtwmodel}
In this model (appendix of 
\cite{dtw}), the messengers in complete $SU(5)$ multiplets
\footnote{Unlike the previous model, this model
preserves gauge coupling unification.} 
couple to a SUSY breaking field which is an {\em adjoint} of $SU(5)$
rather than a singlet whereas the supersymmetric mass of the messengers
comes from an $SU(5)$ singlet field. This modifies the mass spectrum for
the gauginos and scalars. 

The superpotential is:
\begin{eqnarray}
W & = & \lambda _3 X q_1 \bar{q}_1 + \lambda _2 X l_1 \bar{l}_1 +
X \xi ^2 \nonumber \\
 & & + 
\lambda ^{\prime}_3 S ( q_1 \bar{q}_2 + 
q_2 \bar{q}_1 )  + \lambda ^{\prime} _2 S (l_1 \bar{l}_2 +
l_2 \bar{l}_1 ),
\end{eqnarray}
where $X$ is an a adjoint of $SU(5)$, $S$ is a gauge singlet
and $q$, $l$ form a 
${\bf 5}$ of $SU(5)$.
With $F_X \neq 0$, $F_S \sim 0$ and $\lambda^{\prime} S \gg \lambda X$
\footnote{With this superpotential, the vev's of $X$ and $S$ are
undetermined at tree-level, but with a modified superpotential
and/or radiative corrections, the hierarchy of the vev's can be achieved.},
the $R$-symmetry breaking scale is $\sim X$
\footnote{Under the $R$-symmetry which is {\em not} broken by the
vev of $S$, but {\em is} broken by the vev of $X$, the $R$-charges are
$(X, S, l_1, \bar{l}_1, l_2, \bar{l}_2) = 
(2, 0, x, -x, 2 + x, 2 - x)$.} and
we get the following mass spectrum when the messengers are integrated 
out (assuming $\sqrt{F_X} \ll S$)
\footnote{This model also has ``gauge'' messengers:
the gauge multiplet corresponding to the broken generators of $SU(5)$ 
has a non-supersymmetric spectrum (due to $F_X \neq 0$ and $X$ being
an adjoint) and couples to the MSSM
sparticles. However, the contribution of 
the ``gauge'' messengers to the soft masses is $\sim \alpha / (4 \pi)
F_X / M_{GUT}$ which is much smaller than the contribution of the
$q$, $l$ messengers which is $\sim  \alpha / (4 \pi)
F_X / S$ (assuming $S \stackrel{<}{\sim} 1/10 \; M_{GUT}$).}:
\begin{eqnarray}
m^2_{\hbox{scalar} \; i} & \approx & 2 \; \left( \frac{F_X}{S} \right) ^2 
\Biggl[ C_3^i \; \left( \frac{\lambda _3}{\lambda _3^{\prime}}
\right) ^2 \; 
\left( \frac{\alpha _3}{4 \pi} \right) ^2 
+ C_2^i \; \left( \frac{\lambda _2}{\lambda _2^{\prime}}
\right) ^2 \; \left( \frac{\alpha _2}{4 \pi} \right) ^2 \Biggr. \nonumber \\
 & & + \Biggl. \frac{3}{5} \; Y_i^2 \; \left( \frac{\alpha _1}{4 \pi} 
\right) ^2 \left( \frac{2}{5} \; \left( \frac{\lambda _3}
{\lambda _3^{\prime}}
\right) ^2 + \frac{3}{5} \; \left( \frac{\lambda _2}
{\lambda _2^{\prime}}
\right) ^2 \right) \Biggr],
\end{eqnarray}
\begin{eqnarray}
M_1 & \sim & \frac{\alpha _1}{4 \pi} \; 
\frac{F_X \; X}{S^2} \; \Biggl[ \frac{2}{5}
\frac{\lambda _3^2}{\lambda _3^{\prime \; 2}} 
+ \frac{3}{5} \frac{\lambda _2^2}
{\lambda _2^{\prime \; 2}} \Biggr], \nonumber \\
M_2 & \sim & \frac{\alpha _2}{4 \pi} \; 
\frac{F_X \; X}{S^2} \; \frac{\lambda _2^2}
{\lambda _2^{\prime \; 2}}, \nonumber \\
M_3 & \sim & \frac{\alpha _3}{4 \pi} \; 
\frac{F_X \; X}{S^2} \; \frac{\lambda _3^2}{\lambda _3^{\prime \; 2}},
\end{eqnarray}
where the gauge and the Yukawa couplings are evaluated 
at the messenger scale $\sim \lambda^{\prime} S$.
Since the field $X$ is an $SU(5)$ adjoint whereas
$S$ is a singlet, we get
\begin{eqnarray}
\frac{\lambda _2}{\lambda _3} & = & - \frac{3}{2}, \nonumber \\
\lambda ^{\prime} _2 & = & \lambda ^{\prime} _3
\end{eqnarray}
at the GUT scale.
If we assume that $\lambda$, $\lambda^{\prime} \ll g_A$ (where
$g_A$'s are the gauge couplings), then $\lambda _3 / \lambda _3^{\prime}$
and $\lambda _2 / \lambda _2^{\prime}$ are RG invariant
(below the GUT scale).
This implies that $\left( \lambda _2 / \lambda _2^{\prime} \right) 
\left( \lambda _3^{\prime} / \lambda _3 \right) \approx -3/2$ at the
messenger scale
{\em also} so that the mass spectrum at the messenger scale is:
\begin{eqnarray}
m^2_{\hbox{scalar} \; i} & \approx & 2 \; \Lambda ^2 \;
\Biggl[ C_3^i \; 
\left( \frac{\alpha _3}{4 \pi} \right) ^2 + C_2^i \; \frac{9}{4} \;
\left( \frac{\alpha _2}{4 \pi} \right) ^2 + \frac{21}{20} \; Y_i^2 \; 
\left( \frac{\alpha _1}{4 \pi} 
\right) ^2 \Biggr],
\end{eqnarray}
\begin{eqnarray}
M_1 & \sim 
& \frac{\alpha _1}{4 \pi} \; \Lambda \; x \; \frac{7}{4}, \nonumber \\
M_2 & \sim 
& \frac{\alpha _2}{4 \pi}  \; \Lambda \; x \; \frac{9}{4}, \nonumber \\
M_3 & \sim & \frac{\alpha _3}{4 \pi}  \; \Lambda \; x,
\label{dtwMA}
\end{eqnarray}
where $\Lambda \equiv \lambda _3 F_X / (\lambda _3^{\prime} S)$
and $x \equiv (\lambda _3 X) / (\lambda _3^{\prime} S )$.
We see that the gaugino masses can be suppressed relative 
to the scalar masses due
to $R$-symmetry breaking scale being smaller than the messenger scale,
{\it i.e.}, by a factor $\sim \lambda X / ( \lambda ^{\prime} S)$. Thus,
there is an independent parameter for the gaugino masses so that
gauginos can be sub-TeV even if the scalars are multi-TeV.
Also, the SUSY {\em breaking} mass for messenger leptons is
larger (due to $\lambda _2 > \lambda _3$)
than that for the quarks so that $y_{actual}$ is larger.

Thus, the soft SUSY breaking masses can be written in terms
of {\em two} 
parameters which we choose to be the gluino mass at the messenger
scale, $M_3 (0)$, and the right-handed up-squark mass at the 
messenger scale, $m_{\tilde{u}^c} (0)$ \footnote{
The fundamental parameters are $M_3 (0)$, $m_{\tilde{u}^c} (0)$,
$\lambda _t$, $\mu$, $B \mu$ and $M_{mess}$ and the {\em free} (or input)
parameters
are chosen to be $M_3 (0)$, $m_{\tilde{u}^c} (0)$, $\tan \beta$ and
$M_{mess}$.
As before,
$\mu$ and $B \mu$ are determined from the minimization
conditions.}. 

\begin{figure}
\centerline{\epsfxsize=1\textwidth \epsfbox{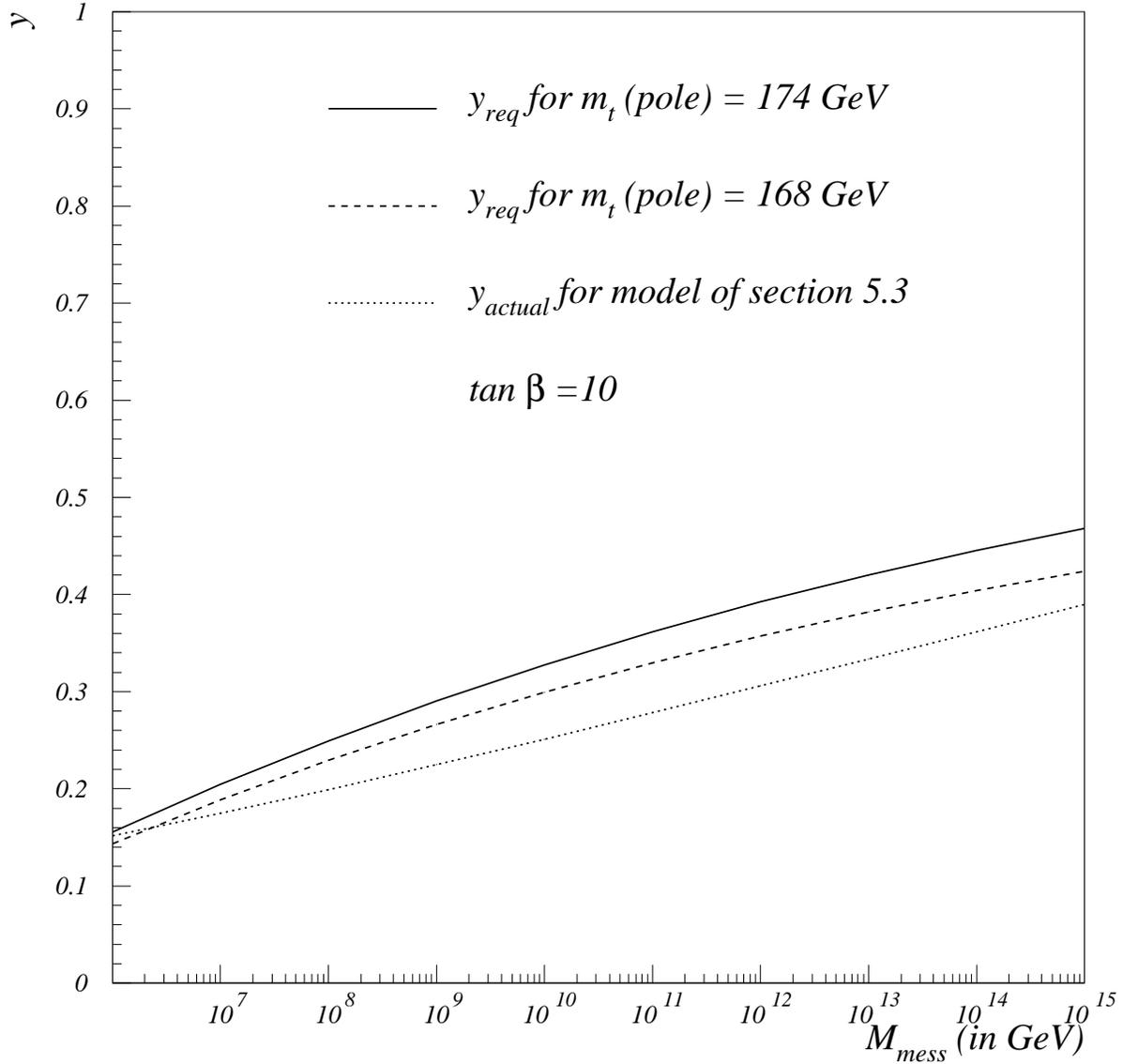}}
\caption{The ratio of the Higgs mass to the stop mass (Eq. (\ref{yactualdef}))
in the Dimopoulos,
Thomas, Wells model (section \protect\ref{dtwmodel}) (dotted line)
as a function
of messenger scale. The solid and the dashed lines are
the ratios of
Higgs mass to the stop mass {\em required} for focussing (Eq. (\ref{yreq}))
for $m_t$ (pole) $ = 174$ GeV
and $168$ GeV, respectively, with 
$\tan \beta =10$ and
the renormalization scale $\mu _{RG} \approx 2$
TeV.
}
\protect\label{ydtw}
\end{figure}

In Fig.\ref{ydtw}, we plot $y_{actual}$ (which depends only
on $M_{mess}$ for a given model: see Eq.
(\ref{yactualdef})) for this model as a function
of the messenger scale and also $y_{req}$
for $\tan \beta =10$ and (since
we want stops to be multi-TeV) 
the renormalization scale $\mu _{RG} \approx 2$
TeV. 
We see that $y_{actual} \approx y_{req}$, especially
for smaller values of $m_t$ and very low or high
messenger scales, so that we
expect focussing in this model (certainly, focussing is ``better''
than in minimal GM: compare Figs.\ref{ygmsb1} and \ref{ydtw}) .

\begin{figure}
\centerline{\epsfxsize=0.8\textwidth \epsfbox{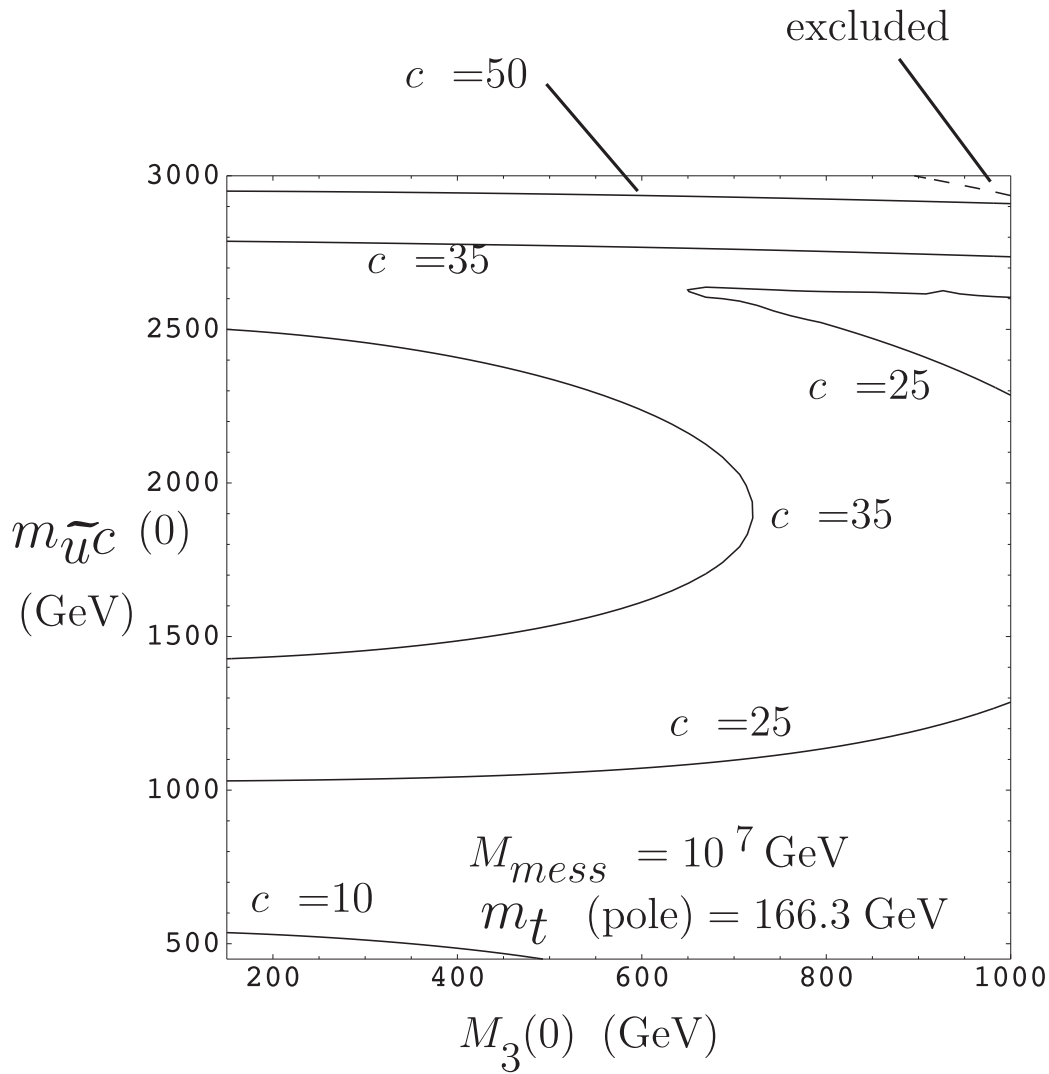}}
\vspace{-0.2in}
\caption{The fine-tuning 
$c$ (Eq. (\protect\ref{c})) (solid lines)
in the Dimopoulos, Thomas, Wells model
(section \protect\ref{dtwmodel}) as a function of the right-handed up-squark
mass ($m_{\tilde{u}^c}(0)$) and the gluino mass ($M_3(0)$) at the
messenger scale.
The messenger scale is $\approx
10^7$ GeV, $\tan \beta =10$ 
and the top quark (pole) mass is $166.3$ GeV.
The tiny region above the dashed line in the upper right corner
has chargino mass
smaller than $90$ GeV or $\mu^2 < 0$ and is therefore excluded.}
\protect\label{dtwmucM30}
\end{figure}

\begin{figure}
\centerline{\epsfxsize=1\textwidth \epsfbox{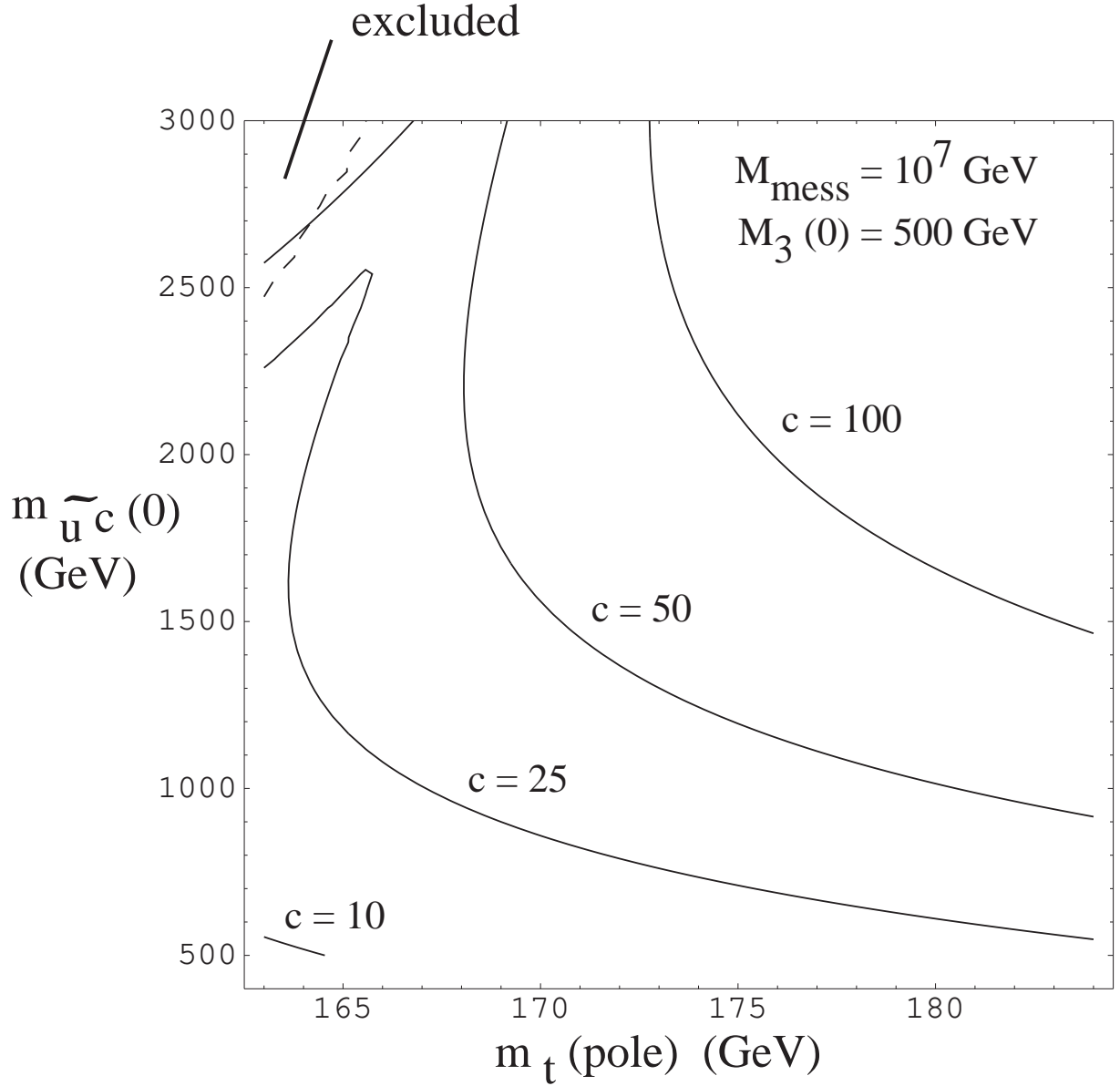}}
\caption{The fine-tuning $c$ (Eq. (\protect\ref{c})) (solid lines)
in the Dimopoulos, Thomas, Wells model
(section \protect\ref{dtwmodel}) as a function of the right-handed up-squark
mass at the messenger scale ($m_{\tilde{u}^c} (0)$) 
and the top quark (pole) mass.
The messenger scale is $\approx
10^7$ GeV, $\tan \beta =10$ and the gluino
mass at the messenger scale, $M_3 (0)$, is $500$ GeV.
The region to the left of 
the dashed line (the upper left corner) has chargino mass
smaller than $90$ GeV or $\mu^2 < 0$ and is therefore excluded. 
}
\protect\label{dtwmucmt}
\end{figure}

\begin{figure}
\centerline{\epsfxsize=1\textwidth \epsfbox{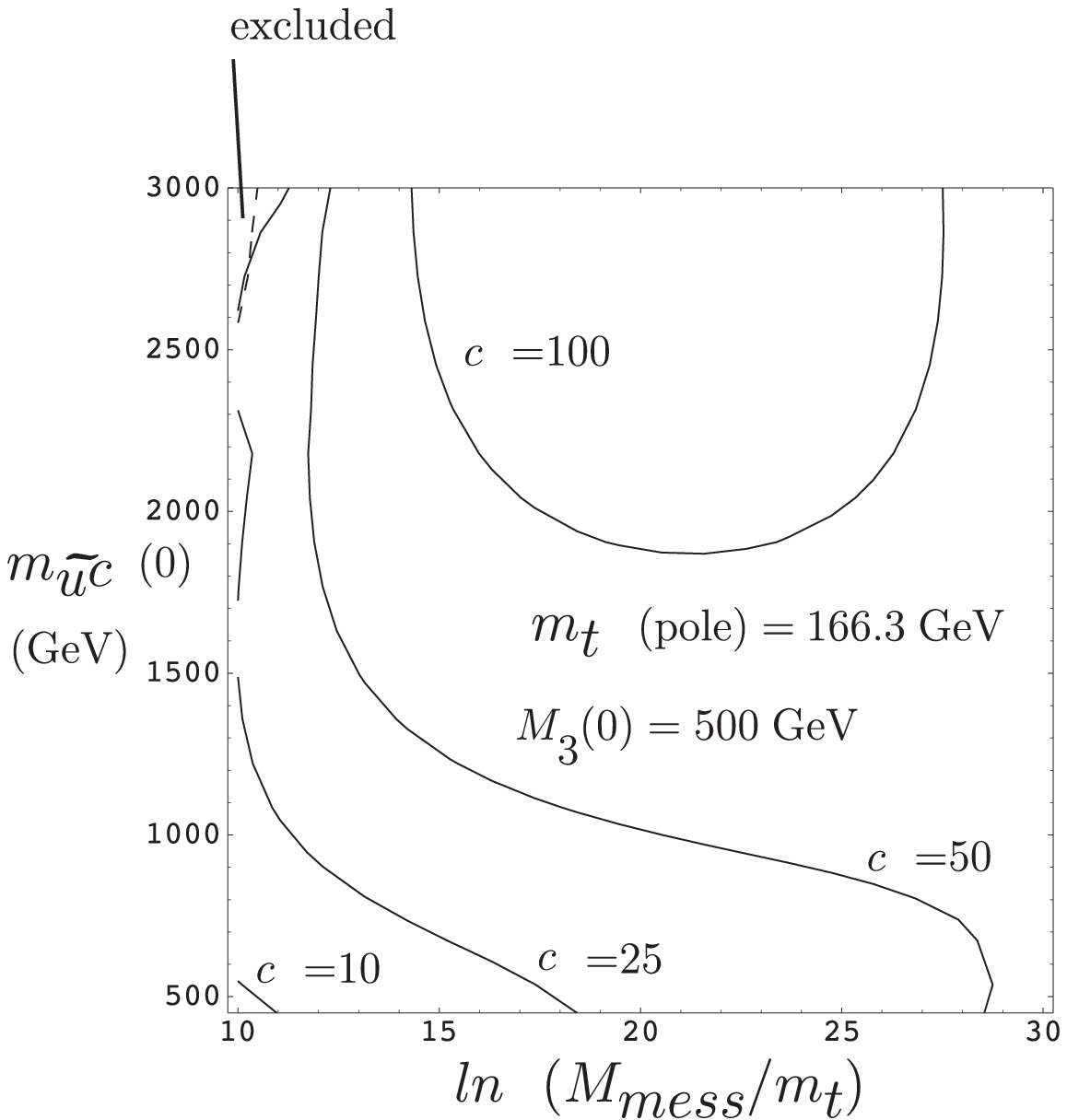}}
\caption{The fine-tuning $c$ (Eq. (\protect\ref{c})) (solid lines)
in the Dimopoulos, Thomas, Wells model
(section \protect\ref{dtwmodel}) as a function of the right-handed up-squark
mass at the messenger scale
($m_{\tilde{u}^c} (0)$) and the messenger scale, $M_{mess}$. The gluino
mass at the messenger scale, $M_3 (0)$, is $500$ GeV, $\tan \beta =10$ and 
the top quark (pole) mass is $166.3$ GeV.
$\ln (M_{mess} / m_t) = 10$ corresponds to $M_{mess} \approx
4 \times 10^6$ GeV and $\ln (M_{mess} / m_t) = 30$ corresponds to
$M_{mess} \approx 2 \times 10^{15}$ GeV.
The tiny region to the left of the dashed line (upper left corner)
has chargino mass
smaller than $90$ GeV or $\mu^2 < 0$ and is therefore excluded.}
\protect\label{dtwt0muc}
\end{figure}

In Figs.\ref{dtwmucM30}, \ref{dtwmucmt} and \ref{dtwt0muc}, 
we plot the contours of fine-tuning for this model.
The renormalization scale, $\mu _{RG}$, is chosen to be
$m_{\tilde{u}^c}(0) \approx \sqrt{m_{\tilde{t}_1} m_{\tilde{t}_2}}$ (where
$m_{\tilde{t}_{1,2}}$ are the physical stop masses).
As mentioned above
and as can be seen in Fig.\ref{dtwmucmt}, 
focussing ($y_{actual} \approx y_{req}$)
is ``better'' for smaller values of $m_t$. So, in Figs.\ref{dtwmucM30}
and \ref{dtwt0muc}
we use $m_t$ (pole) 
$ = 166.3$ GeV, which is $\sim 1 \sigma$ below the central value.

The right-handed squark (other than stop)
masses at the weak scale are $\approx
m_{\tilde{u}^c}(0)$ whereas the left-handed squarks (including
the stop) are (slightly) heavier. The right-handed stop mass 
(at the weak scale) is typically (depending
on $M_{mess}$)
$(1/2 \; -\;1) \times m_{\tilde{u}^c}(0)$
(due to the effect of the top quark Yukawa coupling and the larger
value of $m_{H_u}^2 (0)$ than in minimal GM). The right-handed
slepton mass is $\sim (1/4 \; - \; 3/4) \times m_{\tilde{u}^c}(0)$
(again, depending on $M_{mess}$). Thus,
there is a lower limit of $\sim 500$ GeV
on $m_{\tilde{u}^c} (0)$ from collider searches for right-handed
sleptons and stops.
Gaugino masses can be obtained (in terms
of $M_3(0)$)
from Eq. (\ref{dtwMA}) using 
$M_A(t) \propto \alpha _A (t)$ (at one-loop). 

We see from the figures that 
even for $m_{\tilde{u}^c} (0) \sim 1 \; - \; 3$
TeV, it is possible to have $c \stackrel{<}{\sim} 100$
for a wide range of $M_{mess}$ and for {\em smaller}
values of $m_t$ ($\sim 1 \sigma$ below the central value). 
Also, for these values of $m_t$
and either very low
($\sim 10^7$ GeV) or high ($\sim
10^{15}$ GeV) messenger scales, some values of
$m_{\tilde{u}^c} (0)$ in the range 
$\sim 1 \; - \; 3$
TeV are almost as natural as sub-TeV $m_{\tilde{u}^c} (0)$
\footnote{
For example, we see in Fig.\ref{dtwmucmt}
that for $m_t$ about $1 \sigma$ below the central value,
the fine-tuning contours are sort of parallel to the $m_{\tilde{u}^c}$
axis. However, for larger values 
of $m_t$ and for intermediate $M_{mess}$ ($\sim 10^{12}$ GeV), since
focussing is not so good (see Fig.\ref{ydtw}),
{\em sub}-TeV $m_{\tilde{u}^c} (0)$ is {\em more
natural} than $m_{\tilde{u}^c} (0) \stackrel{>}{\sim}1$ TeV, even
though it is still possible to have
$c \stackrel{<}{\sim} 100$ for
$m_{\tilde{u}^c} (0) \stackrel{>}{\sim}1$ TeV and these values
of $m_t$ and $M_{mess}$.}. 
Thus, multi-TeV ($\sim
1 \; - \; 3$ TeV)
squarks (including left-handed stop) can be natural in this model. If
$m_{\tilde{u}^c}(0) \stackrel{<}{\sim} 
1.5$ TeV, then the right-handed slepton
can be discovered at a TeV lepton collider.
The right-handed stop can be 
lighter than $1$ TeV if $m_{\tilde{u}^c}(0) \stackrel{<}{\sim} 2$ TeV and so
is likely to be detected at the LHC whereas the other squarks might escape 
detection at the LHC 
(at least with $10$ (fb)$^{-1}$ of integrated luminosity)
if $m_{\tilde{u}^c}(0) \stackrel{>}{\sim} 1$ TeV.

Since there is an independent parameter
for the gaugino masses, it is possible that
all the gauginos will be detected at the LHC/TeV
lepton collider provided
$M_3 (0) \stackrel{<}{\sim} 500$ GeV. From Eq. (\ref{dtwMA}), we
see that the wino mass is enhanced 
as compared to minimal GM 
({\it i.e.}, $M_2 / M_3 \sim 1$ at the weak scale)
thereby reducing the gaugino mass contribution
to the Higgs (mass)$^2$ (as discussed earlier)
so that $c \stackrel{<}{\sim} 100$
is possible even for $M_3 (0) \stackrel{>}{\sim} 500$ GeV 
(see Fig.\ref{dtwmucM30}). In fact, for some values of
the other parameters, 
$M_3 (0) \stackrel{>}{\sim}500$ GeV is {\em as natural as} $M_3 (0) 
\stackrel{<}{\sim} 500$ GeV.
In this case, depending on 
$M_{mess}$, we can have wino and gluino masses at the weak scale
$\sim 1$ TeV or heavier
(without
any fine-tuning)
and thus the wino will be beyond the reach of
a TeV linear collider and the detection of gluino at the LHC might 
require {\em more than} $10$ (fb)$^{-1}$ of integrated luminosity. 
As before, for the region
of the parameter space with $c \stackrel{<}{\sim} 100$, we have
$\mu \stackrel{<}{\sim} 600$ GeV so that the Higgsinos 
(to be precise the
neutralino and chargino which have Higgsinos as the
dominant component) can be discovered at
a ($1$ or $1.5$) TeV lepton collider also
(or at the LHC, if $\mu$ is very small). 
Since $M_1 \sim 1/3 \; M_3$ (at the
weak scale) in this model, the neutralino 
which has bino as the dominant component
is (typically) the lightest SM superpartner.

As mentioned earlier, for smaller values of $\tan \beta$ (say less
than $5$ and for a fixed $m_t$)
$y_{req}$ is larger so that ``focussing'' will not be effective
in this model \footnote{Also, in this model, 
multi-TeV squarks implies $m_{H_d}^2 \sim$
(TeV)$^2$ leading to fine-tuning for small $\tan \beta$.}. 
However,
the above results are also valid for $5 \stackrel{<}{\sim}
\tan \beta \stackrel{<}{\sim}20$ (since the top quark Yukawa coupling
is roughly the same and the bottom quark Yukawa coupling
is small in this range
of $\tan \beta$).

The spectrum in the above two GMSB models (sections \ref{qlmodel}
and \ref{dtwmodel})
(with multi-TeV squarks)
\footnote{We have not been very precise about
the mass spectrum and detectability of sparticles at colliders
in these two models.
A detailed analysis (including two-loop
RGE's) of fine-tuning and mass spectrum (phenomenology)
in these models is in progress.} 
should be contrasted to the spectrum in supergravity mediation
with multi-TeV 
{\em universal} scalar mass (and focussing so that multi-TeV
stops are natural) \cite{feng}.
In this supergravity model, in addition to squarks, 
(right-handed)
slepton masses are {\em also} multi-TeV and hence
will escape detection at TeV lepton colliders
(whereas, as mentioned
earlier, in
GMSB models right-handed sleptons are lighter than squarks
by a factor $\sim \alpha _1 / \alpha _3$) and gaugino masses
are {\em sub}-TeV so that there is no fine-tuning (whereas in the models
of sections \ref{qlmodel} and \ref{dtwmodel}, gluino mass can
be $\sim 1$ TeV or heavier without resulting in fine-tuning).
Also, for $M_{mess} \stackrel{<}{\sim} 10^7$ GeV in
the models of sections \ref{qlmodel}
and \ref{dtwmodel},
the intrinsic scale of SUSY breaking, $\sqrt{F}$, can be less than
$\sim 10^3$ TeV ($F_X / M_{mess} \sim
100$ TeV to give $\sim$ TeV squarks). In this case, it is possible (unlike
in supergravity mediation)
that the 
lightest SM superpartner, {\it i.e.}, 
the next-to-lightest
supersymmetric
particle (NLSP) (the gravitino is LSP if $M_{mess} \ll M_{GUT}$)
will decay {\em inside} 
the detector (after being produced directly or in cascade 
decays). The NLSP
decays give the well-studied $\not \! \! p _T + \gamma \gamma$ (or
$ZZ$) (if
the NLSP is a neutralino)
and $\not \! \! p _T + 2$ leptons (if the NLSP is a right-handed
slepton) signals of GM.

{\em Minimal} GM with the SUSY breaking scale $m_{SUSY} \sim
\alpha / ( 4 \pi) \; \Lambda \stackrel{>}{\sim}1$ TeV can also have
multi-TeV squarks (and gluino) and possibly a wino mass
$\sim 1$ TeV and a {\em sub}-TeV (right-handed)
slepton so that its signals might mimic those in the models of 
sections \ref{qlmodel}
and \ref{dtwmodel}. However, due to focussing, the models of 
sections \ref{qlmodel}
and \ref{dtwmodel} have $\mu \stackrel{<}{\sim} 500$ GeV 
(and hence are not very fine-tuned) whereas 
the minimal GM model with multi-TeV squarks will have $\mu \stackrel{>}
{\sim} 1$ TeV (and thus is severely fine-tuned)
so that Higgsinos can be detected (at a
TeV linear collider) in the former models
unlike in the latter \footnote{Similarly, supergravity models with
multi-TeV squarks, but {\em without} focussing, will have $\mu \stackrel{>}
{\sim} 1$ TeV and hence the Higgsinos will be beyond the reach of a TeV
linear collider.}.

In the ``more-minimal'' supersymmetric models \cite{moreminimal}, 
the first and second generation squarks and sleptons (inlcuding right-handed
sleptons, {\it i.e.}, complete ${\bf \bar{5}}$ and ${\bf 10}$
of $SU(5)$) are heavy, say about $10$ TeV so
that the supersymmetric flavor problem is (partly)
alleviated
whereas
the stops (and possibly the
sbottom and stau) are lighter than $\sim 1$ TeV 
for naturalness ({\it i.e.},
to keep $\mu \stackrel{<}{\sim}500$ GeV). So, the first and second
generation squarks and sleptons are beyond the reach of LHC/TeV 
linear colliders while the top squarks
should be detected at the LHC in these models \footnote{Of course,
the stops can also be multi-TeV in which case the model will be
severely fine-tuned with $\mu \stackrel{>}{\sim}1$ TeV
so that ``Higgsinos'' will be beyond the reach of TeV linear colliders.}.
Thus, the collider signals of these ``more-minimal''
models are different from the models of sections \ref{qlmodel}
and \ref{dtwmodel} which have top (and other)
squarks $\sim 1 \; - \; 3$ TeV
(which might escape detection at the LHC) 
without resulting in
fine-tuning, {\it i.e.}, keeping $\mu \stackrel{<}{\sim}500$ GeV
and also (in some cases) 
{\em all} right-handed slepton masses $\sim 500$ GeV.

\section{Summary}
In this paper, we have analysed whether in GMSB, the
soft SUSY breaking mass spectrum
can be such that the (up-type)
Higgs (mass)$^2$ (evaluated at the weak scale) is {\em in}sensitive
(or weakly sensitive) to the scalar (Higgs
and stop) masses at the messenger scale ({\it i.e.},
the boundary scalar masses)
-- this has been called ``focussing'' in
\cite{feng} where {\em supergravity} mediation
was considered. Then, the stop masses (and also 
other squark masses since they are related) can be multi-TeV {\em without}
leading to $m_{H_u}^2 \sim - (\hbox{TeV})^2$ at the weak scale
(and the consequent fine-tuning in electroweak symmetry breaking). 

In minimal GMSB, the
Higgs mass at the boundary is too small compared to the stop masses
so that this focussing does not happen. Also in minimal GM, the gaugino
(especially the gluino) masses will also be multi-TeV if the
scalars are multi-TeV, resulting in fine-tuning of $m_Z$.

Thus, in order for multi-TeV squarks to be natural 
in (non-minimal) 
models of GM, we require
a) additional contribution to the Higgs mass (or a non-minimal
mesenger sector) so that the ratio of the Higgs mass to the stop mass
at the boundary can be higher (to achieve focussing)
and b) suppression of gaugino masses
(or a larger wino mass relative to gluino mass) so that even if
the scalar masses are multi-TeV, the gaugino masses do not result in 
fine-tuning. 

We discussed (general)
ideas to 
satisfy these two requisites
and also studied concrete models of GM which have multi-TeV ($
\sim 1 \;
- \; 3$ TeV) squarks
without leading to fine-tuning in EWSB. 
In some cases, these models
have (in addition to $\sim 1 \;
- \; 3$ TeV squarks) wino and gluino with a mass of $\sim 1$ TeV 
(without fine-tuning) and
right-handed slepton mass $\sim 500$ GeV (so that
right-handed sleptons can be detected at TeV lepton
colliders) and (for very low messenger
scales) lightest SM superpartner which decays {\em inside} the detector.
Thus, the collider signals of these models can be
different from the minimal supergravity models with multi-TeV squarks
(and no fine-tuning) \cite{feng} which 
have {\em multi}-TeV {\em sleptons} (which will escape detection at
TeV lepton colliders) 
and {\em sub}-TeV gauginos (and of course a stable 
lightest SM superpartner) 
so that it is possible to distinguish (experimentally)
between
the two scenarios. Unlike minimal GM models with multi-TeV squarks
(which have $\mu \stackrel{>}{\sim}1$ TeV), 
the (non-minimal) models studied here can have $\mu \stackrel{<}{\sim}
500$ GeV so that the Higgsinos (to be precise the
neutralino and chargino which have Higgsinos as the
dominant component) can be detected at a TeV linear collider.

\end{document}